\documentclass[10pt,sigconf]{acmart}
\usepackage{graphicx}
\graphicspath{{figures/}}
\newtheorem{definition}{Definition}
\usepackage[linesnumbered,ruled]{algorithm2e}
\usepackage[colorinlistoftodos]{todonotes}
\usepackage{listings}% http://ctan.org/pkg/listings
\AtBeginDocument{%
  \providecommand\BibTeX{{%
    \normalfont B\kern-0.5em{\scshape i\kern-0.25em b}\kern-0.8em\TeX}}}

\setcopyright{acmcopyright}
\copyrightyear{2018}
\acmYear{2018}
\acmDOI{10.1145/1122445.1122456}

\acmConference[Woodstock '18]{Woodstock '18: ACM Symposium on Neural
  Gaze Detection}{June 03--05, 2018}{Woodstock, NY}
\acmBooktitle{Woodstock '18: ACM Symposium on Neural Gaze Detection,
  June 03--05, 2018, Woodstock, NY}
\acmPrice{15.00}
\acmISBN{978-1-4503-9999-9/18/06}

\begin{document}

\title{ML-AQP: Query-Driven Approximate Query Processing based on Machine Learning}
\author{Fotis Savva}
\affiliation{%
  \institution{University of Glasgow}
  \city{Glasgow}
  \country{UK}}
\email{f.savva.1@research.gla.ac.uk}

\author{Christos Anagnostopoulos}
\affiliation{%
  \institution{University of Glasgow}
  \city{Glasgow}
  \country{UK}}
\email{christos.anagnostopoulos@glasgow.ac.uk}

\author{Peter Triantafillou}
\affiliation{%
  \institution{University of Warwick}
  \city{Warwick}
  \country{UK}}
\email{p.triantafillou@rwarwick.ac.uk}

\begin{abstract}
As more and more organizations rely on data-driven decision making, large-scale analytics become increasingly important. However, an analyst is often stuck waiting for an exact result. As such, organizations turn to Cloud providers that have infrastructure for efficiently analyzing large quantities of data. But, with increasing costs, organizations have to optimize their usage. Having a cheap alternative that provides speed and efficiency will go a long way. 
Concretely, we offer a solution that can provide approximate answers to aggregate queries, relying on Machine Learning (ML), which is able to work alongside Cloud systems. Our developed lightweight ML-led system 
can be stored on an analyst's local machine or deployed as a service to instantly answer analytic queries, having low response times and monetary/computational costs and energy footprint. To accomplish this we leverage the knowledge obtained by 
previously answered queries and build ML models that can estimate the result of new queries in an efficient and inexpensive manner. The capabilities of our system
are demonstrated using extensive evaluation with both real and synthetic datasets/workloads and well known benchmarks.
\end{abstract}

\maketitle




\section{Introduction}
\begin{figure}
\begin{center}
\includegraphics[height=4.5cm,width=5cm,keepaspectratio]{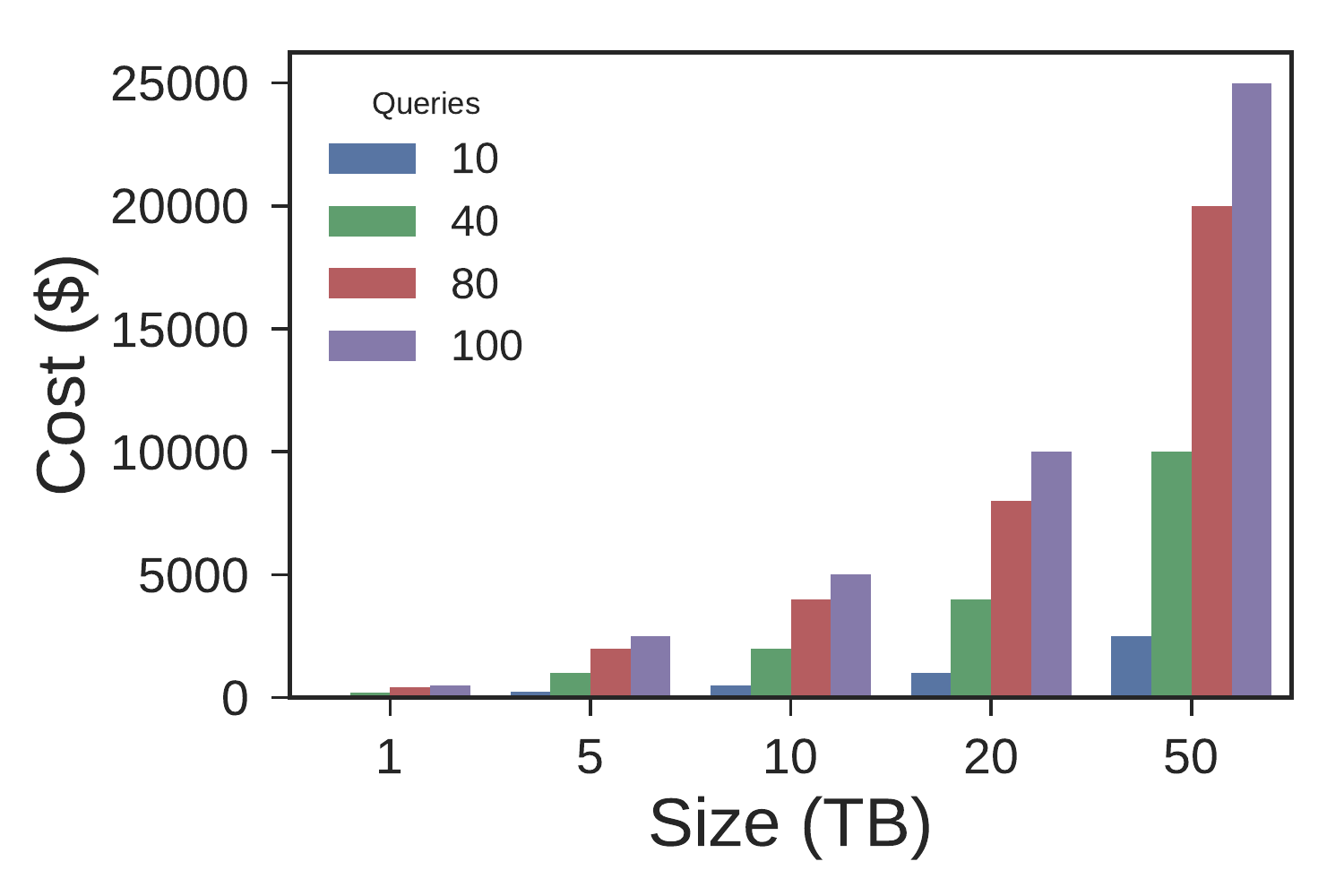}
\caption{Costs associated with using cloud-managed databases (BigQuery). The x-axis is the amount of data used per query and the y-axis is the associated costs with the average number of queries daily.}
\label{fig:clouds-costs}
\end{center}
\end{figure}
%Data growing exponentially - BD engines/DBs no longer enough - Additional constraints for interactivity by exploratory analysis processes - exploratory analysis often conducted by extracting statistics from data with queries that have large selectivity and then a much narrower focus - AQP engines are not lightweight they require huge samples - In need of complimentary solutions when interactivity is a must - AQP engines BD engines require high resources often reside in expensive cloud environments - Need something to be in the middle to allow for (1) efficient computation of statistics for exploratory analysis (2) lightweight in terms of storage so that they can reside on a local machine and not in the cloud (3) accurate in terms of answer - Data access is expensive AQP engines relying on data are not lightweight on the other hand queries are ignored but can often provide enough information to estimate statistics.  ( Have previously been used to both guide indexing schemes or other data structures facilitating aggregate estimation. Most recently they've been used along with ML Models to compute aggregates such as COUNT/AVG and Regression Queries. ) 
With the rapid explosion of data volume and the adoption of data-driven decision making, organizations are struggling to analyze data efficiently and inexpensively. Hence, a lot of companies turn to Cloud providers that maintain large-scale data warehouses \cite{ sato2012bigquery, gupta2015amazon} 
able to store and process large quantities of data. But, the problem still remains: as  
multiple queries are issued by multiple analysts, clusters are often overburdened and costs rise significantly. 
For instance, looking at Figure \ref{fig:clouds-costs}, we can observe the exponential costs 
associated with an increasing data size. The associated cost (in y-axis) is obtained after 
multiplying the cost of scanning an amount of data (in x-axis) with varying number of queries shown as colored bars \footnote{Data For this graph are obtained from \url{https://cloud.google.com/bigquery/pricing}}. 
Most importantly, during data analysis, it is of high importance to be able to extract information 
without significant delays so as to not break the \textit{interactivity constraint}, 
which is set around 500ms \cite{liu2014effects}. This constraint supports that any 
answers returned over that limit can have negative effects on an analyst's experience and productivity.
Concretely, analysts often engage in what is called \textit{exploratory analysis} \cite{exploration} 
in order to better understand the data. Such analyses are an invariable step in the process of further constructing hypotheses or constructing and training predictive and inferential models to answer business questions. 
A key to these analyses is that an approximate answer to any query is often enough to move forward. 
As such, over the last few decades research has focused into systems that allow Approximate Query Processing (AQP)\cite{garofalakis2001approximate,park2017databaseaqp,agarwal2013blinkdbaqp,kandula2016quickraqp} to facilitate the process 
of data analysis with this in mind. 
By trading off some of the accuracy they allow for order of magnitude speed-ups in execution. 

Although AQP systems offer a straight-forward and efficient solution to the problem of efficiently processing queries, 
they come at a cost. They require large samples and would have to reside in the same Cloud system which makes them costly to maintain as every operation carries a cost. 
In addition, in cases where multiple analysts are using the Cloud system or AQP engine to process queries, they might be stuck waiting for their job to execute as multiple operations could be in the queue.
Hence, what we propose is a \textit{complementary system} to that of AQP engines
addressing some of their shortcomings and better highlighting their strengths.  

We envision a system that is computationally lightweight and can be stored in an analyst's machine or in a central server, away from the main backend analytic system of choice. 
This allows for the exploratory process to be executed locally at the analyst's machine, 
without overburdening the cloud system, thus, saving up resources (and money) to be used in cases where accurate answers are needed. 
From the cloud provider's standpoint, our solution could act as a \textit{pseudo}-caching mechanism 
to reduce load when it is necessary thus allowing for other operations to run.

What makes such a system possible, and the salient feature of our approach, is the availability of a number of previously executed queries (in log files). 
Leveraging previously executed 
queries allow for the creation of Machine Learning (ML) models that 
can estimate the results of new unseen queries. As exploratory analysis is often made 
up of operations that filter the data and then return descriptive statistics (aggregates) on the resulting subsets, 
we can construct ML models that predict such statistics. 
An ML model can effectively learn to associate the input parameter values of a query with the obtained result. 
Subsequent predictions can then be made in milliseconds, thus, 
fulfilling the interactivity constraint. In addition, most ML models are orders of magnitude lighter and can be stored on any device. 
%The analysts can use those trained models, \textit{locally}, to answer any queries they might have with a trade-off in accuracy. 

The desiderata are: Derived models are lightweight (in terms of memory footprint), easy to configure and fast to train, supporting error guarantees for their predictions, able to deal with updates to both query patterns and DB updates, able to handle all types of aggregate functions (AFs), and ensure high accuracy.
Lastly, such mechanism 
does not require access to any of the data neither at training time, as is the case with current sampling-based and ML-based AQP approaches, nor at prediction time as 
what is being queried is the \textit{actual} trained model. 
Concretely, our technical contributions are as follows:
\begin{itemize}
    \item A flexible vectorized representation for (SQL) queries, to be used by ML models;
%    \item A light-weight aggregate query result estimation with low \textbf{storage}, \textbf{computational}, \textbf{monetary} costs.
    \item The first AQP engine (ML-AQP) that mines query logs (query-driven) and develops ML models meeting all above desiderata;
    \item Up to 5 orders of magnitude greater efficiency than the state of the art sampling-based techniques;%, with small memory footprint and small preprocessing (model training) time;
    \item A new method for providing probabilistic error guarantees, based on Quantile Regression to complement approximate answers;
    \item Support for all AFs (including MIN/MAX which cannot be supported by current approaches);
    \item A comprehensive performance evaluation using synthetic and several real-world data sets and workloads which substantiate performance claims.
\end{itemize}{}
\section{Preliminaries and Supported Queries}
%Figure for viewing DB as black-box
We abstract every (back-end) Cloud analytics system (both relational and non-relational) as a black box. Thus, essentially, queries are regarded as executing over sets of multi-dimensional points \footnote{A single row in a table can be considered as a multi-dimensional point} to which a number of operations are performed to return a result. Both \textit{non-relational} and \textit{relational} databases can be considered as large collections of attributes either grouped in a collection of normalized tables or being part of a single de-normalized data set. 
We can store our data in either of the two settings and the result of a query will still be the same. 
% Figure \ref{fig:storing-data} shows an example of this, in which data are stored under different formats without any difference to the results produced. 
It is just the way of performing data manipulation and aggregation that differs. 
In the remainder of this section, we demonstrate how common operations in a relational schema can be performed using our proposed representation. This is without loss of generality to any kind of data-storage system and it is merely used as the abstraction should be familiar to the reader.
Given the above, our first concern is to develop an appropriate representation for queries so that an ML model can associate the query representation with the query results and learn to predict answers for 'similar' queries. 

% \begin{figure}
% \begin{center}
% \includegraphics[height=4.5cm,width=5cm,keepaspectratio]{db-storage}
% \caption{Data stored in different formats have no effect on the result returned by a query}
% \label{fig:storing-data}
% \end{center}
% \end{figure}
\begin{definition}(Data \& Attributes)
As we are unaware of the underlying data storage format, 
we adopt a generic assumption of all data being collections of attributes, 
where a dataset $\mathcal{B}$ is a collection of real-valued $d$-dimensional vectors $\mathbf{a} = (a_1,\ldots, a_d)$, 
such that $\mathcal{B} = \{\mathbf{a}\}_{i=1}^n$. The $\mathbf{a}$ vector holds values for $d$ attributes.
\end{definition}

\begin{definition}(Aggregate Functions)
Aggregate Functions (AF) are applied to the returned result-set and 
mapping a set of returned values to a scalar result $y\in \mathbb{R}$. 
An AF can be applied to a specific attribute; AFs commonly include functions such as \texttt{COUNT}, \texttt{AVG}, \texttt{SUM}. They are typically used in an SQL-style query along with various predicates and joins.
\end{definition}

\begin{definition} (Predicates)
Predicates are used to restrict the number of rows (data vectors) returned by a query. 
Predicates can be considered as a sequence of negations, conjunctions, and disjunctions ($\neg,\lor, \land$) over attributes with equality and/or inequality constraints ($\leq, \geq, =$). 
A well known predicate is the \textit{range-predicate}. 
A \textit{range-predicate} effectively restricts an attribute $a_i$ 
to be within a given range [$lb$, $ub$] with $a_i \geq lb \land a_i \leq ub$. 
To effectively model a sequence of predicates, we assign two \textit{meta-attributes} 
for each attribute $a_i$ and consider every predicate as a range-predicate. 
The two meta-attributes are equal to the [$lb$, $ub$] of a range-predicate. 
For instance, without loss of generality, assume the three following predicates applied 
on a dataset with a single attribute $a_1$: (1) $a_1 \geq lb$, (2) $a_1 = c$, where $c$ is a numerical value, and (3) $a_1 \geq lb \land a_1 \leq ub$. We construct two meta-attributes for each case as follows: (1) $(a_{1,lb}, -)$, where $-$ could be set to \texttt{NULL} and $a_{1,lb}$ is the supplied $lb$ value, (2) $(a_{1,c},a_{1,c})$, where $a_{1,c}=c$, and (3) $(a_{1,lb}, a_{1,ub})$.
\label{def:predicate}
\end{definition}

% \subsection{Support of Operators in Predicates}
% Using range-predicates provided in Definition \ref{def:predicate}, 
% a wide range of queries can be supported as shown in the example. 
% The only thing remaining is to distinguish between various operators such 
% as $(\leq, <, \geq, >, =)$ and various sequences of such predicates using conjunction, disjunction, and negation. 
% First of all, it is straightforward to derive appropriate value to transform various equality operators to more general ones such as transforming $<$ to $\leq$, etc. 
% We simply decrease the number on the right-hand side by one $x < 10 \to x \leq 9$. 
% As for the sequence of negations, conjunctions, and disjunctions, 
% we rely on enough queries being present for the ML training algorithm 
% to be able to distinguish between such operators. 
% As this is beyond the scope of our current work, we left it in our future agenda.
%Structural information incorporation of another meta attribute to represent conjunction/disjunction/negation between predicates. If existence of multi-predicates on same attribute, we could find way to estimate which one is more important and leave that one. We should not worry about queries making no sense and we should also understand that the general purpose of this project is to increase efficiency on most common queries and not outlier queries appearing once.
\subsection{Transforming Aggregate Queries to Vectors}
\subsubsection{SPA Queries}
\label{def:queries}
We first consider \textit{Selection-Projection-Aggregate} (SPA) queries, in which a single aggregate 
is the result of a query; that is made up of a single relation and multiple predicates. 
Given our definition of predicates, we obtain a meta-vector which is made up of all 
the constraints $\mathbf{m} = (a_{1,lb}, a_{1,ub},\ldots, a_{d,lb}, a_{d,ub})$ across all attributes. 
Hence, each SPA query can be represented by a meta-vector $\mathbf{m}\in \mathbb{R}^{2d}$. For all attributes that are part of the data set but not part of the query we leave the values of their associated meta-attributes as \texttt{NULL}. For instance, a simple SPA aggregate query is the following  applied over a data set $\mathcal{B}$ with attributes $\mathbf{a} = (a_1, a_2, a_3)$:
\begin{lstlisting}
    SELECT AF($a_i$)
    FROM $\mathcal{B}$
    WHERE $a_1\geq x_1 \land a_2\leq x_2$
\end{lstlisting}{}
The meta-vector is $\mathbf{m}= (x_1, \texttt{NULL} , \texttt{NULL}, x_2, \texttt{NULL}, \texttt{NULL})$.

\subsubsection{SPJA Queries}
To effectively model \textit{SPJA} (Selection-Projection-Join-Aggregate) queries we first redefine what it means to join two or more tables together from a query representation perspective.   
We assume an architecture in schema design where, if multiple tables exist, then these tables 
are made up of a large \textit{fact} table along with much smaller \textit{dimension} tables. 
This is widely accepted in the literature \cite{park2017databaseaqp,agarwal2013blinkdbaqp,hall2016trading}. Specifically, in a designed AQP system by Google \cite{hall2016trading} it is mentioned that all queried relations are pre-joined so that JOINs are not performed at query runtime. 
As a result, a data analyst simply queries the large fact table using \textit{equi}-joins whenever they wish 
to project more attributes to the result set. Therefore, the dimensionality of the initial row (data vector) obtained from the \textit{fact} table is simply increased. 
However, it is evident that the result set is still only affected by the predicates in the selection. Assuming these kind of JOINs and that the number of rows is not affected by the resulting JOIN our initial representation is appropriate without any changes.

\subsubsection{GROUP-BY Queries}%Figure
\begin{figure}[!htbp]
\begin{center}
\includegraphics[height=6.5cm,width=7cm, keepaspectratio]{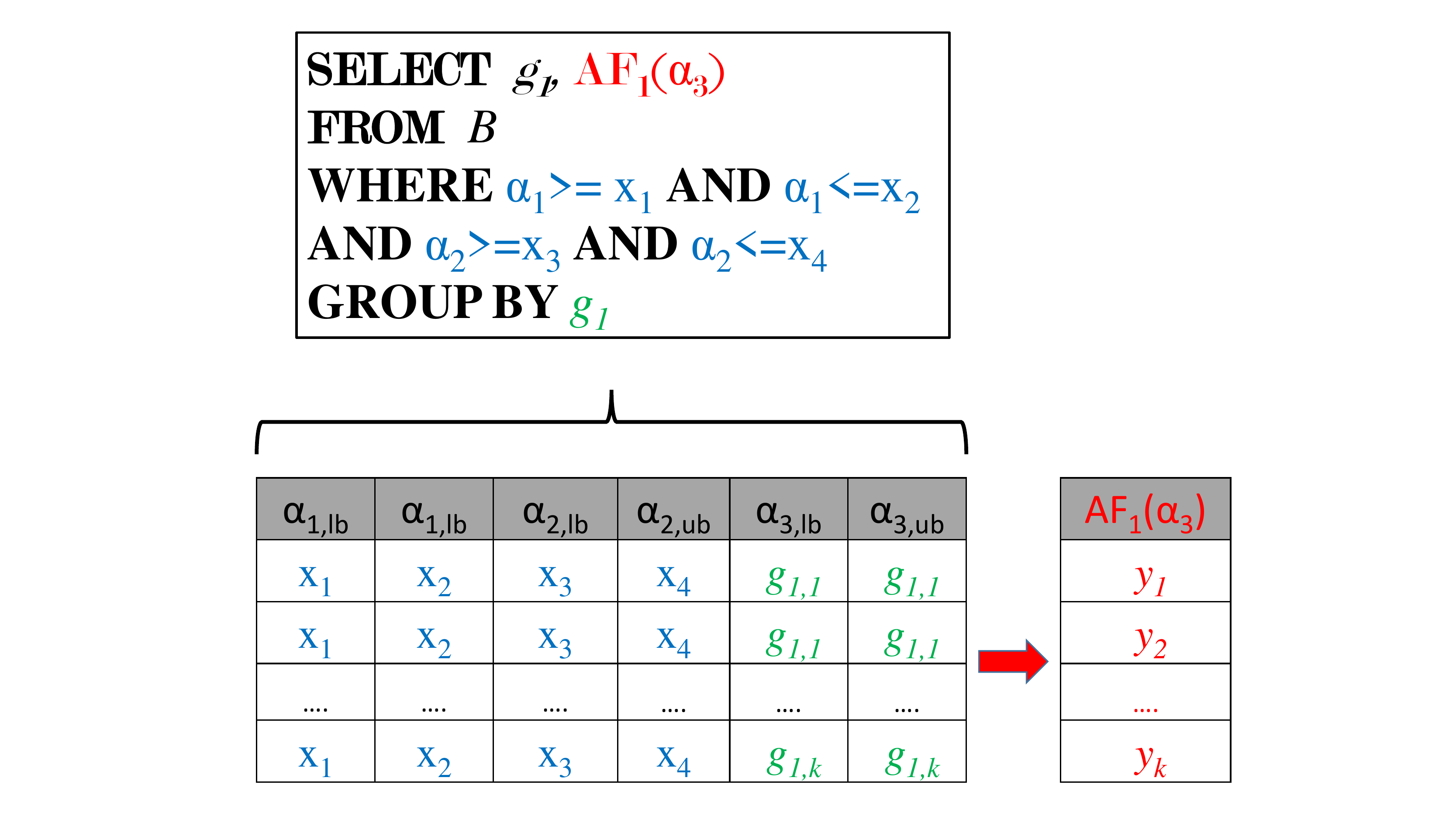}
\caption{How to vectorize \texttt{GROUP-BY} queries.}
\label{fig:groupby-handling}
\end{center}
\end{figure}

Supporting \texttt{GROUP-BY} queries is crucial for data analytics, as analysts often issue queries to explore differences between cohorts via grouping. 
A \texttt{GROUP-BY} query is an application of the same AF onto different sub-groups defined by an attribute. 
Hence, for an attribute $a_i$, we use a \texttt{DISTINCT} operator to identify the
different groups and subsequently use the vectorization process of an SPA query on the individual groups. The result of the operator would be a set of group values $\mathcal{G} = (G_1,\ldots, G_k)$, which can be used as an equality predicate to construct $k$ different queries. An example of this is shown at Figure \ref{fig:groupby-handling}. An SQL query with a \texttt{GROUP-BY} clause is issued and the colored parts of this query are extracted. Suppose that the group-by attribute used is $a_3$, $a_3=g_1$. The predicate values $(x_1, x_2, x_3, x_4)$ and the extracted group values $(G_1, \ldots, G_k)$ are used to construct $k$ meta-vectors in which the values for $(a_{1,lb}, a_{1,ub}, a_{2,lb}, a_{2,ub}$ contain the same values for all rows as the filter-predicates are applied for each group value $G_{1,i}$. The last two columns $a_{3,lb},a_{3,ub}$ are used to store the group values. Each one of those meta-vectors will become associated with the output of the corresponding AF.
This is similar to the formulation of Database Learning \cite{park2017databaseaqp}. However, we do not limit the number of generated queries to $1000$ as suggested by the formulation of \cite{park2017databaseaqp}. Hence, an arbitrarily large number of groups can be supported by our formulation. 

\subsubsection{Handling Categorical Attributes}
\label{sec:categorical}
Some attributes might hold categorical values instead of numerical. 
An accepted approach is to restrict the length of the categorical attributes 
to the currently longest of each categorical attribute  \cite{kraska2018case}. Another option is to construct various \textit{dummy} columns each one denoting 
a value included in the categorical attribute $a_i$\cite{kuhn2013applied}. Suppose $N$ distinct values for $a_1 = (A_1, \ldots, A_N)$, then $N$ dummy columns are created, with its rows having a value of $\{0,1\}$, that being a mapping $A_i \to \{0,1\}$.
However, the inherent  problem with this option is the explosion in dimensionality of 
the query vector $\mathbf{m}$ as its dimensionalty now becomes $2d+N$. To this end, an effective encoding scheme can 
be an injective function, such as various hash functions, that provide 
an effective mapping from a categorical attribute to a real number, i.e., 
$\mathcal{A} \mapsto \mathbb{R}$. 
In our implementation for ML-AQP we use a combination of both. 
The first technique is used for attributes with low cardinality $<1000$. For tree-based ML algorithms, 
this has been shown to work best \cite{kuhn2013applied}. The latter technique is used for attributes with high cardinality $>1000$. 

\subsection{Overall Support for Queries \& Limitations}
Overall, with this representation we are able to support a large fraction 
of the aggregate queries commonly in an OLAP setting, 
from simple multi-predicate aggregation queries to queries that include JOINs and GROUP-BYs. We can provide support for foreign-key joins as this is the case for multiple AQP engines \cite{park2017databaseaqp}. Specifically our setting does not make any assumptions as to what type of aggregate functions are used. To ML-AQP the response variable is a scalar $y$, associated with a meta-vector $\mathbf{m}$. Subsequently it tries to identify patterns in $\mathbf{m}$ that would allow it to predict a future $y_\text{new}$ when given an $\mathbf{m}_\text{new}$. Therefore it is agnostic to the AFs used. This is in contrast to most sampling based AQP engines \cite{agarwal2013blinkdbaqp} which restrict the number of aggregates supported. 
In addition, in the presence of textual filters (\texttt{LIKE '\%product'}) the same approach to other categorical attributes can be applied. Meaning that the pattern is considered as a string and encoded following the approach described at Section \ref{sec:categorical}. 
For JOINs which do not simply extend the dimensionality but instead introduce less/more tuples in the result we do not explicitly represent them in the current meta-vector. As we described, usually such schema designs are avoided when conducting analyses over large amounts of data \cite{hall2016trading}.
In addition, derived attributes for GROUP-BYs cannot be supported with our current formulation and instead such queries have to be partially executed to obtain the derived attributes. 

% \subsection{Aggregate Estimation and ML Models}
% Once a vectorized representation for queries is obtained, we use ML algorithms to associate the meta-vectors with their results. 
% An aggregate result is a scalar value $y \in \mathbb{R}$, thus, 
%  a dataset $\mathcal{C} = {(\mathbf{m}_i, y_i)}_{i=1}^n$ is derived from past query executions and their results. The goal of any statistical learning algorithm will then be to \textit{minimize} the expected 
% loss (difference or discrepancy) 
% between the \textit{true} query result of $y$ and an estimated value of it, $\hat{y}$, derived from the trained ML model. As each AF has a different underlying conditional distribution, we train ML models for each of those AFs. 
% Concretely, AFs such as \texttt{COUNT}, \texttt{SUM}, \texttt{AVG} will each be associated with specific ML model(s). Additional ML models are created for those AFs that refer to specific attributes. 
% Although the number of ML models seems to be increasing their storage footprint, as we will examine in our experimental section it is minimal compared to the storage requirements of a sampling-based approach. It is important to note that we do not a-priori construct models for each possible AF or combination of AF-attribute. Instead, a model is trained for an AF, if and only if there are queries in the query log referring to this specific AF. Implicitly, ML-AQP builds models over AFs that are frequently used. This significantly reduces the number of models that have to be built.
\section{System Architecture}
\begin{figure}[!htbp]
\begin{center}
\includegraphics[height=5cm,width=7cm, keepaspectratio]{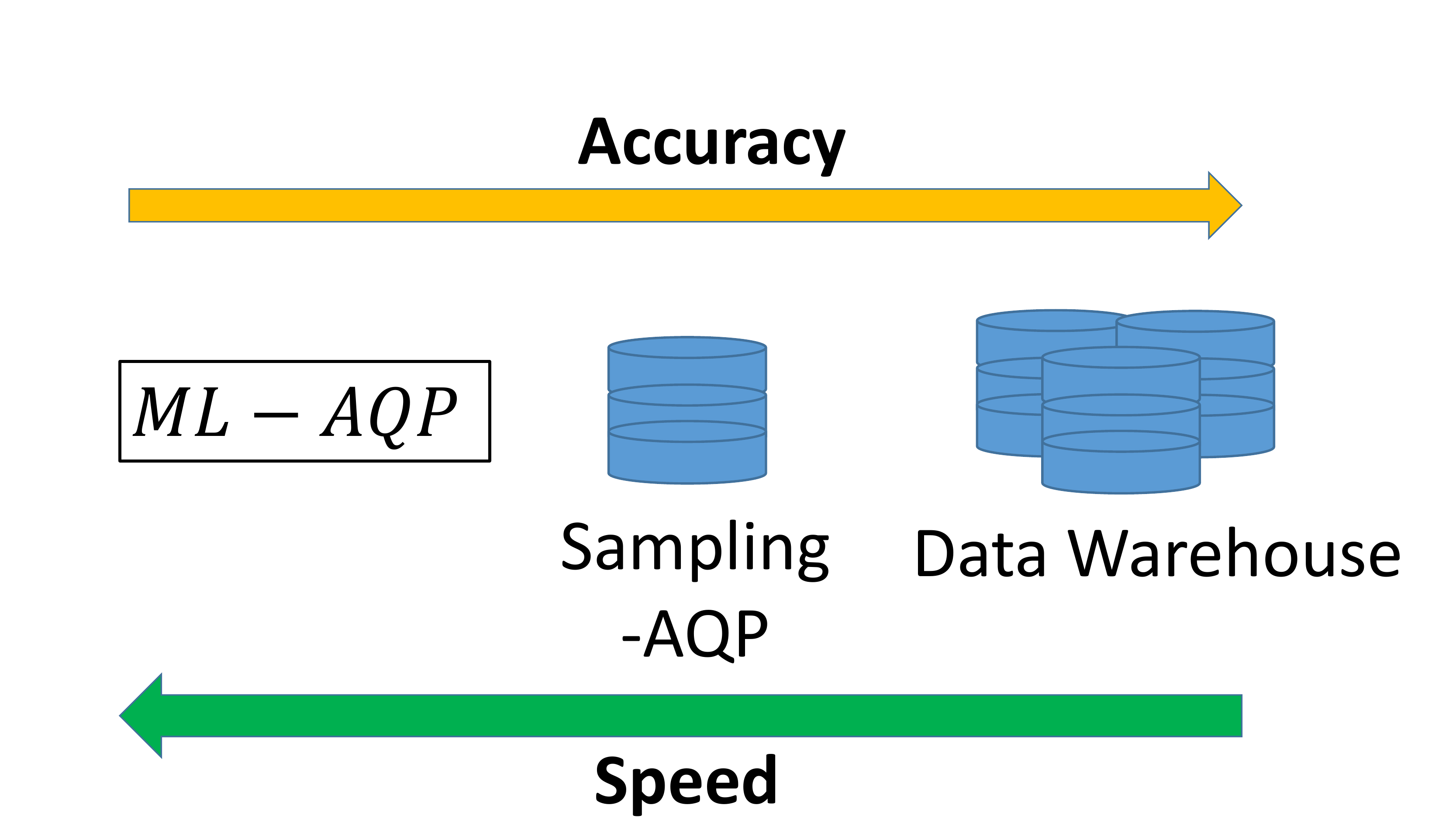}
\caption{The ML-AQP within the complete data analytics stack. Starting from ML-AQP, analysts can choose a 
system going from left to right, if they require more accuracy. If speed is essential, 
they can choose from right to left.}
\label{fig:system-placement}
\end{center}
\end{figure}
Figure \ref{fig:system-placement} shows holistically how ML-AQP complements 
the data analytics stack. Our system sits between 
cloud-based data warehouses and/or \textit{Sampling}-based AQP engines. 
Using all three components (ML-AQP, sampling-based AQP engines, data warehouses), the analyst can choose which one to use based 
on their needs of efficiency and accuracy. 
A system could also make this choice based on the resources available. 
Hence, if a cloud system is experiencing heavy loads, it could direct queries 
to either the sampling-based AQP (S-AQP) engine or ML-AQP. 
A useful analogy is to think of each of the three components as 
the \textit{cache}, \textit{RAM} and \textit{Disk} components of a computer. 
Caches and RAMs can often not hold the data required (hence the lack of accuracy), 
but the disk will always hold the true answer. 
However, this comes at a cost in efficiency. Therefore in our case, ML-AQP can act as the cache of the data analytics stack, sampling-based AQP engines as the RAM and finally cloud-based engines as the disk. 
% \begin{figure}[!htbp]
% \begin{center}
% \includegraphics[height=5cm,width=7cm, keepaspectratio]{figures/maqp-modules.pdf}
% \caption{ML-AQP system architecture}
% \label{fig:system-arch}
% \end{center}
% \end{figure}

To better explain the overall architecture followed in ML-AQP, 
we explain each component as part of two distinct modes: 
(a) \textit{Training} mode and (b) \textit{Prediction or Production} mode.
%All of the sub-components of ML-AQP are shown at Figure \ref{fig:system-arch}.
During \textit{Training} mode, queries are either executed at the Data 
Warehouse or the S-AQP and become associated with their results. We can also utilise pre-computed queries stored at log files.  
ML-AQP leverages those queries to build training sets of query-result pairs for the ML models. 
Training these ML models transitions ML-AQP to the \textit{Prediction/Production} mode to which 
queries are now being transformed into the described vectorial representation $\mathbf{m}$ 
and their results are estimated by ML models. 

\begin{figure}[!htbp]
\begin{center}
\includegraphics[height=5cm,width=7cm, keepaspectratio]{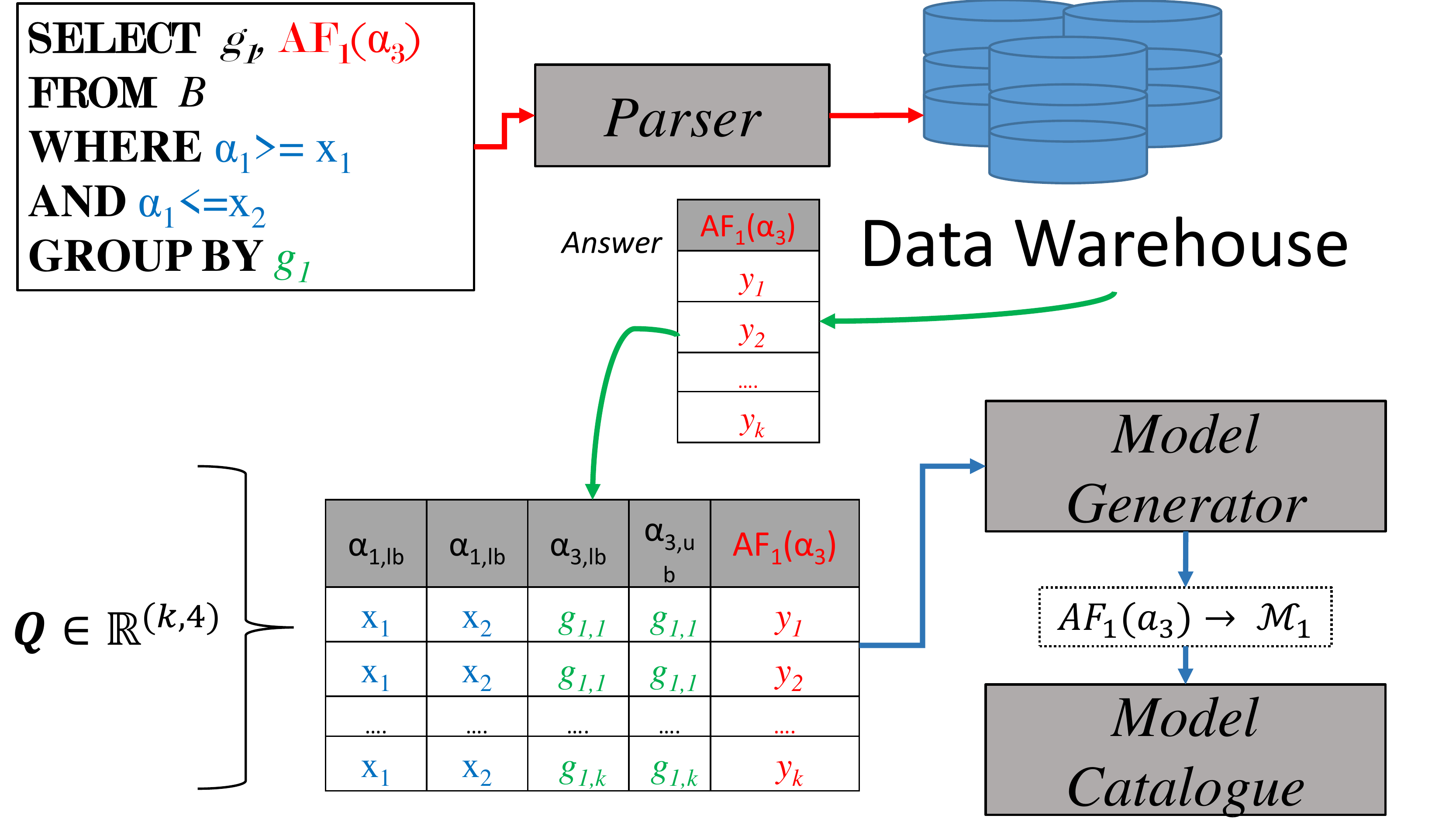}
\caption{ML-AQP during Training}
\label{fig:system-training}
\end{center}
\end{figure}
The complete flow and interaction of the components, is shown at Figure \ref{fig:system-training}. Initially, at \textit{Training} mode, each query is parsed, through the Parser, 
and the projected AFs are extracted $f_1,\ldots, f_n$, along with the included 
predicates $p_1,\ldots, p_m$ and any GROUP-BY attributes $g_{1}, \ldots, g_{k}$.
In the example at Figure \ref{fig:system-training}, the extracted AF is  $f_1 = \texttt{AF}_1(a_3)$, the resulting predicates are $p_1= (a_1 \geq x_1)$ and $p_2 = (a_1 \leq x_2)$ and the GROUP-BY attribute is $g_1$. 
For the predicates, we construct an $\mathbf{m}\in \mathbb{R}^{2d}$ meta-vector, 
where $d$ is total number of attributes in a data set.
Each meta-vector is associated with a number of results $y_1, \ldots, y_n$, 
obtained from the executed AFs. If no GROUP-BY attributes are included then we call this a single query $\mathbf{q} = (\mathbf{m}, \mathbf{y}), \mathbf{q}\in \mathbb{R}^{2d+n}$. For any, GROUP-BY attribute $g_{i}$ a \texttt{SELECT-DISTINCT} query is executed and its result is cached in GROUP-BY catalogue $D$. The catalogue $D$ is a mapping from the \texttt{GROUP-BY} attribute $g_{i}$ to its set of distinct values $D: g_{i} \to \mathcal{G}_{g_{i}}$.  Caching its result allows their reuse during \textit{Prediction/Production} mode. For any \texttt{GROUP-BY} attributes that are frequently used together such as $(g_i,g_j)$ we also cache their set of distinct values.
Given the values returned for $g_{i}$, $\mathcal{G}_{g^{(i)}} = \{G_{g_{i},1}, \ldots, G_{g_{i},|G_{g_{i}}|}\}$, 
we construct multiple single-queries which have different results for $y_1, \ldots, y_k$ as they correspond to different groups. Hence, in the case of \texttt{GROUP-BY} queries, a single query has a matrix representation 
for its meta-vectors and their associated results $\mathbf{Q} = (\mathbf{M},\mathbf{Y}), \mathbf{Q}\in \mathbb{R}^{ (|G_{g_{(i)} }|)\times(2d+n)}$, this result is shown at Figure \ref{fig:system-training}. The same procedure occurs for each executed query, which results in the collection of possibly sparse vectors (as not each attribute is often included in the predicates or the GROUP-BY clause). 
Because of this, we store all of the processed queries in a sparse matrix to save space. 
Once we finish parsing and constructing the representation for each query in our training set, 
we use the Model Generator to construct/train models $\mathcal{M}_i, \mathcal{M}_n$ for each AF encountered and an association is created $\texttt{AF}_1(a_3) \to \mathcal{M}_1$. 
The models are then serialized and stored in a \textit{Model Catalogue}.

\begin{figure}[!htbp]
\begin{center}
\includegraphics[height=5cm,width=7cm, keepaspectratio]{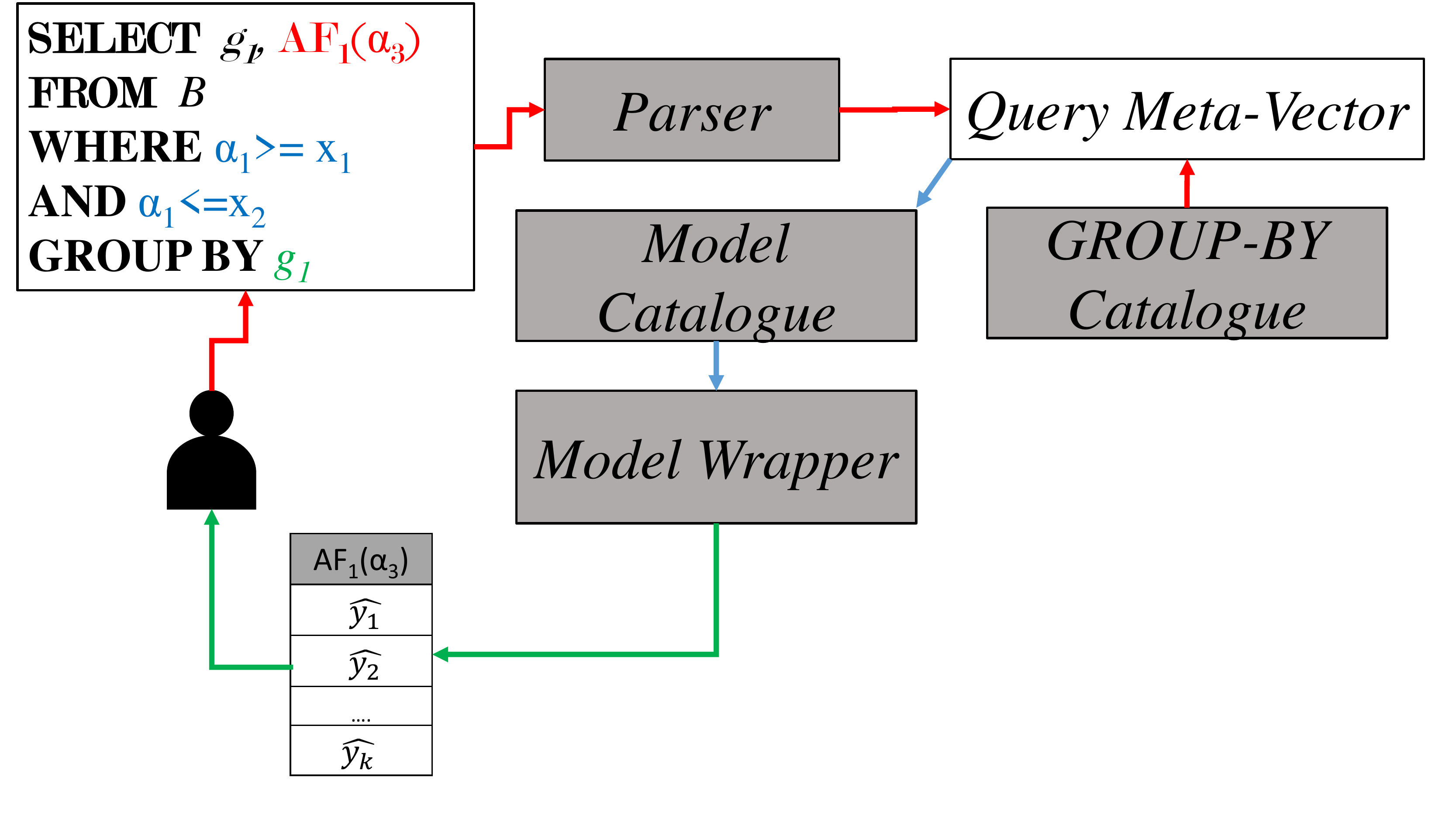}
\caption{ML-AQP during Production/Prediction}
\label{fig:system-production}
\end{center}
\end{figure}
A similar process occurs during the \textit{Prediction/Production} mode when a new (SQL) query is issued at the ML-AQP. The complete flow of interactions between sub-components is shown at Figure \ref{fig:system-production}. During initialization of the ML-AQP, both the GROUP-BY catalogue and Model Catalogue are loaded in memory. Consider the same query used as an example at Figure \ref{fig:system-training}, but this time the query is not sent for execution but instead ML-AQP is asked to predict its answer. Again the Parser is used to extract the same elements (predicates, AFs, GROUP-BY attributes). A vectorized representation of the query is constructed at \textit{Query Meta-Vector}. If a \texttt{GROUP-BY} statement exists, the resulting meta-vector is a matrix $\mathbf{M}$ and the values for $g_i, \ldots, g_k$ are obtained from the \texttt{GROUP-BY} catalogue storing different $\mathcal{G}_{g_i}$. If the group-by attributes have been cached together then that result is fetched from the catalogue. 
The necessary AFs to be estimated are identified and their models are fetched from Model Catalogue. The Model Wrapper is used to query the model and estimate the results given the meta-vector $\mathbf{m}$. The result(s) are then returned to the user in an efficient manner as no data are accessed and the only overhead is the inference time of a model.
\section{Machine Learning Specifics}
Each query result  $y$ from the training pair $(\mathbf{m},y)$ derives from an unknown true function $f(\cdot)$. 
Such function produces answers w.r.t an unknown conditional distribution $p(y|\mathbf{m})$. 
Our aim is to approximate the \textit{true} functions $f$ for each aggregate function e.g., \texttt{COUNT}, \texttt{AVG}, \texttt{MIN}, \texttt{MAX}, \texttt{SUM}, etc. Supervised ML algorithms are adopted to minimize the expected prediction loss between the actual answer $y = f(\mathbf{m})$ from the true function $f(\cdot)$ and the predicted $\hat{y}$ from an approximated function $\hat{f}(\cdot)$ generated by the ML model: 
\begin{equation}
\label{eq:obj-error}
\min_{f} \mathbb{E}[\mathcal{L}(f(\mathbf{m}), \hat{f}(\mathbf{m}))]
\end{equation}
The loss $\mathcal{L}(\cdot, \cdot)$ can be usually measured using \textit{absolute} or \textit{squared} loss functions:
$
\mathcal{L} = |f(\mathbf{m}) - \hat{f}(\mathbf{m})| \mbox{ or } \mathcal{L} = (f(\mathbf{m}) - \hat{f}(\mathbf{m}))^2.
$
Supervised learning algorithms minimize the expected prediction loss using \textit{training} examples, which they use to learn the true function $f$. Such training examples are the
query-answer pairs in $\mathcal{C}$, which are drawn from $P(y|\mathbf{m})$ and are thus a sample of this distribution. Hence, by minimizing 
the \textit{sample} error we expect to obtain good 
estimation over any unseen query. Our objective over the sample shown at \textit{Eq.}\ref{eq:objective} is to minimize the Mean Squared Error (MSE) as it is commonly used in regression algorithms \cite{friedman2001elements}: 
\begin{equation}
\frac{1}{N}\sum_{i}^N(y_{i} - \hat{f}(\mathbf{m}_{i}))^2.
\label{eq:objective}
\end{equation}
Note, various supervised learning algorithms minimize a variant of the objective in (\ref{eq:objective}) including regularization parameters as a technique to avoid \textit{overfitting} \cite{friedman2001elements}.

Going back to our use case, we obtain a set of queries and their responses $\mathcal{C} = \{(\mathbf{m}, \mathbf{y})\}, \mathbf{m}\in \mathbb{R}^{2d}, \mathbf{y} \in \mathbb{R}^{n}$. We treat queries that have a matrix representation $\mathbf{Q}$ as collections of single queries. 
This makes sense as essentially each row in $\mathbf{Q}$ corresponds to a single query if instead of the GROUP-BY attribute a single predicate existed restricting $g_i$ to a single value. The task is to train ML models $\mathcal{M}_i,\ldots, \mathcal{M}_n$ that would produce regression functions $\hat{f}_i,\ldots,\hat{f}_n$ that minimize MSE. Multiple such regression algorithms exist and to make the right choice we have to consider some of the properties of the problem at hand.

\subsection{Choice of Machine Learning Models}
A primary concern is that the produced training set $\mathcal{C}$ is inherently sparse as both the parameter vector $\mathbf{m}$ and response vector $\mathbf{y}$ contain NULLs (which can be represented as zeroes in linear algebra) for (a) attributes that do not have any predicates set on them and, hence, $a_{i,lb}, a_{i,ub}$ would be NULL or zero\footnote{A special construct \textit{nan} is placed instead of 0 as 0's are perfectly acceptable values.}, 
(b) queries that do not include an AF producing a response variable $y_i$ would also have NULL values for $y_i$. 
For instance, in a dense matrix representation consider 
two queries calling different AFs; $\texttt{AF}_1$ and $\texttt{AF}_2$, to represent as a matrix we have to set up two columns $\texttt{AF}_1$ and $\texttt{AF}_2$ with the first row (for the first query) having a NULL value for $\texttt{AF}_2$ and the second row having NULL for $\texttt{AF}_1$. To alleviate the NULL response problem we partition the dataset per response/AF such that queries that refer to the same AF are grouped together. However, the input is still sparse. Hence we need algorithms that are able to handle such problems effectively. 

% In addition, we might have to deal with a large number of training queries so we need algorithms that are scalable. 
% For instance, we eliminate the use of Support Vector Regression \cite{smola2004tutorial} as we have found that 
% they cannot handle training examples $>10^4$ if the implementation solves the dual problem in its closed form.

Linear models can often be trained in an online manner using SGD \cite{bottou2012stochastic}, which makes them very efficient. However, they result in simple models which cannot adequately model non-linear relationships without introducing more polynomial terms. 
Models such as Ridge and Lasso \cite{friedman2001elements} regression are interesting variants that include regularization to handle the increased dimensionality of our input. However, we have found that these algorithms do not perform well for our problem. We have also considered the use of Deep Learning, however the models become unnecessarily complex, hard to interpret, expensive to train, difficult to tune and have high inference times \cite{kraska2018case} which could increase the latency of estimating the response of an issued query. hence, they violate the desiderata set earlier.

In light of the above arguments, 
we have made the choice of using Gradient Boosting Machines \cite{friedman2001greedy} (GBM) using efficient-parallel implementations called \texttt{XGBoost}\cite{chen2016xgboost} and \texttt{LightGBM}\cite{ke2017lightgbm}. 
A GBM iteratively fits decision trees, 
at first trying to approximate the response variable and then making this approximation more fine-grained by combining decision trees trained on the negative gradient of the response variable and the produced predictions by the last decision tree. In addition, the highly scalable implementation of GBM by XGBoost and LightGBM allows handling large, high dimensional and sparse input. 

\subsection{Error Guarantees}
A highly desirable feature of any AQP engine is its ability to offer error guarantees to the user, regarding its approximated answers. 
%If the AQP engine provides an estimate $\hat{y}$ for an AF, then the user would like to know whether the \textit{true} value $y$ would be within a (confidence) interval associated with a certain probabilistic guarantee. 
For sampling-based AQP engines, providing such guarantees is relatively straightforward: Using subsampling/bootstraping methods, confidence intervals ($[low, high]$) can be derived, associated with certain confidence levels ($l\%$), indicating that the sampled-population statistic of interest (say, the mean) will fall within the $[low, high]$ range with probability $l\%$ \cite{park2018verdictdbaqp,kandula2016quickraqp, agarwal2013blinkdbaqp, olma2019tasteraqp}. However, this is not appropriate for AQP engines which, instead of inferring population statistics of a sampled population, employ predictive-ML-based models for predicting answers to future questions.

%simple baseline method is to measure the EPE using cross validation
%however this expected error has no guarantees and is constant
%instead use prediction intervals that are similar in nature
%say what they are and their difference from confidence intervals
%how prediction intervals can be computed if assume that y are normally distributed
%how prediction intervals for simple linear regression under assumptions
%bootstrap method to compute better bands. However we need many models
%work on conformal prediction, non-conformity measures (high dimensional input and distance becomes meaningless, not scalable have to store all examples
%jacknife uses LOO so have to fit n models
%end up to Quantile Regression

Therefore, as ML-AQP  relies on predictive/regression ML models, the produced $\hat{y}$ is not an estimator of a population parameter, but a prediction of the answer for a future query (vector $\mathbf{m}$). Hence, we start off our discussion with a naive solution of offering some kind of \textit{error estimation} for aggregate answers. 
All ML models are assessed 
and minimize what is known as the Expected Prediction Error (EPE). This is measured by the loss stated at (\ref{eq:objective}). During training mode, this is an over-optimistic estimate of the \textit{generalizability error} that the specific ML model is associated with. 
The generalizability error is an estimate of the error associated with any future estimations. 
However, because ML models are trained on the set used for measuring EPE, they tend to produce inaccurate estimates. Hence, we use cross-validation \cite{friedman2001elements} that measures the EPE on \textit{out-of-sample} examples that the model did not use during training. Although, techniques such as Leave-One-Out (LOO) and K-Fold \cite{friedman2001elements} produce good estimates for the EPE, this is not probabilistically guaranteed. 
In addition, the EPE is \textit{static} across the input space. This means that, even though an ML model might have learned to predict the answers of certain queries with error $\epsilon_1$ and some others with error $\epsilon_2$, and $\epsilon_1 \ll \epsilon_2$, both sets of queries will have the same EPE associated with them. We find this assumption of a static EPE as undesirable.

Instead of the estimate for EPE, we can use {\textit{Prediction Intervals}}: Unlike confidence intervals, which are used to provide an interval for a population parameter, prediction intervals are used to provide intervals 
that contain the $true$ ($not \ predicted$) value of an aggregate result $y_{n+1}$ of a future query (vector $\mathbf{m_{n+1}}$)). If we knew that the distribution of $y$ is Normal and that any $y$ is independent, prediction intervals could be produced similarly to confidence intervals. Using the sample of $y$ given from the training examples $(y_1, \ldots, y_n)$, we compute the interval: $\overline{y} \pm t_{a}s_n\sqrt{1+\frac{1}{n}}$, where $t_a$ is the $100(1-\frac{a}{2})^{\text{th}}$ percentile of the $t$-distribution, with $\alpha \in (0,1)$ and commonly set to $\alpha=0.01$ or $\alpha=0.05$, $s_n$ is the sample variance of the response variable $y$ and $1-\alpha$ is the \textit{coverage} of the prediction interval.  However, we do not wish to make any parametric assumption about the distribution of $y$. Hence, we resort to other methods outlined below.

As mentioned, bootstrap \cite{bootstrap}, is a prominent method which makes no parametric assumptions about the distribution of $y$. This is used among sampling based AQP engines as well \cite{zeng2014analytical, agarwal2014knowing}. In a sampling-based AQP engine, the bootstrap method is adopted to re-sample the underlying data set $b$ times (where $b$ is usually over $>100$) and produce a distribution of estimates for $y$,  $\hat{y}_{i,1},\ldots, \hat{y}_{i,b}$. 
Let $y_{i,0}$ be the original estimate, with the estimates provided by the bootstrap samples, we then compute the residuals $y_{i,0}-\hat{y}_{i,1}, \ldots, y_{i,0}-\hat{y}_{i,b}$. Using the empirical distribution of residuals, we then compute quantiles, which can be used to produce a confidence interval $[\hat{y}_{i,0} - t_{1-{a/2}}, \hat{y}_{i,0} + t_{a/2}]$. 
Theoretically, ML-AQP could also use the bootstrap method, re-sampling the training dataset $b$ times and constructing $b$ ML models $M_i, \ldots, M_b$. This would yield $b$ estimates for $y$, $\{\hat{y}\}_{i=1}^{b} = \{\hat{f}_i(\mathbf{m})\}_{i=1}^b$ and can similarly produce confidence intervals. However, this methodology would incur the costs of training, maintaining, and predicting the estimates from $b+1$ (if we count the initial model providing the prediction) different ML models. Multiply that by the number of different AFs that need to be learned and the overhead cost of this approach quickly becomes huge.

More recent developments in ML literature focus on building predictive intervals by making use of \textit{conformal inference} \cite{shafer2008tutorialconformal,lei2018distributionconformal, papadopoulos2011regressionconformal}. This technique relies on building a \textit{non-conformity} measure which estimates the difference of two examples i.e., $\mathbf{m}_i$ and $\mathbf{m}_j$. This could be defined as the $L_p$ norm (i.e $p=2$) of the examples $\lVert \mathbf{m}_i-\mathbf{m}_j \rVert_2^2$. But finding the right non-conformity measure in our case is non-trivial as the input vectors are high-dimensional and sparse. 
Distance in this case becomes meaningless \cite{aggarwal2001surprising} and the choice of a \textit{valid} $p$-norm is beyond the scope of this work. In addition, these techniques scan the complete set of previous training examples to find similar and dissimilar examples. 
Thus, all previous queries have to remain stored. This is undesirable, as we would like to discard all of the queries and only deploy models for ML-AQP. 

Therefore, our choice is to employ Quantile Regression (QR) \cite{koenker2001quantile}. %explain why
While typical regression models minimize the EPE and focus on estimating the conditional expectation $\mathbb{E}[y|\mathbf{m}]$, QR tries to estimate the $t^{\text{th}}$ conditional quantile $Q_{y|\mathbf{m}}(t)$. Multiple ML algorithms have been proposed to estimate conditional quantiles \cite{steinwart2011estimating,meinshausen2006quantile, takeuchi2006nonparametric}.
%formally provide the way ath quantiles and how a coverage/prediction interval can be given using quantiles
Formally, given a conditional distribution function for $y$,
\begin{equation}
    F_{y|\mathbf{m}} = \mathbb{P}\{Y\leq y | \mathbf{m}\},
\end{equation}
we define the $t^{\text{th}}$ conditional quantile function as:
\begin{equation}
    q_t(\mathbf{m}) =\inf\{y\in \mathbb{R} : F_{y|\mathbf{m}} > t\}
\label{eq:conditiona-quantile-function}
\end{equation}{}
Where $\inf$ is the \textit{infimum}, which points to $y$ that is less than or equal to all the elements in the defined set. 
Given the conditional quantile function defined in (\ref{eq:conditiona-quantile-function}), we construct prediction intervals using $[q_{t/2}, q_{1-t/2}]$. This defines the lower and upper bounds of the estimated value for $y$ with coverage probability of $100(1-t)\%$.
%how to estimate quantiles using regression using pinball loss
As stated, earlier regression algorithms estimate the conditional expectation of $y$, $\mathbb{E}[y|\mathbf{m}]$ by minimizing EPE. In the same manner, quantile regression algorithms estimate the conditional $t^{\text{th}}$ quantile $Q_{y|\mathbf{m}}(t)$ by minimizing what is known as "\textit{pinball loss}"\cite{koenker2001quantile} :
\begin{equation}
    \rho_t(y, \hat{y}) = 
    \begin{cases}
         t(y-\hat{y}), \quad \text{if}\quad y-\hat{y}>0 \\
         (1-t)(\hat{y}-y), \quad\text{otherwise}
    \end{cases}
\end{equation}
Suppose we have trained two different quantile regression functions $\hat{q}_{t/2}: \mathbf{m}\in \mathbb{R}^{2d} \to Q_{y|\mathbf{m}}(t/2)\in \mathbb{R}$ and $\hat{q}_{1-t/2}: \mathbf{m}\in \mathbb{R}^{2d} \to Q_{y|\mathbf{m}}(1-t/2)\in \mathbb{R}$. Then, a prediction interval for each new query is estimated as: $[\hat{q}_{t/2}(\mathbf{m}_{\text{new}}), \hat{q}_{1-t/2}(\mathbf{m}_{\text{new}})]$ with coverage probability $100(1-t)\%$.

Therefore, ML-AQP provides error guarantees using QR and the statistical tools of prediction intervals and coverage. Specifically, ML-AQP produces a prediction interval $[low, high]$ and a coverage level $l\%$ and guarantees that the $true$ answer to a future query will fall within $[low, high]$ with probability $l\%$.
LightGBM \cite{ke2017lightgbm} offers support for QR and we are also looking into incorporating \cite{romano2019conformalized} to support stricter guarantees.
\section{Updates}
As ML-AQP is deployed, over the course of time there might be \textit{significant} data updates or \textit{drastic} workload shifts that invalidate the patterns learned by the models so far. In general new query patterns or new data might not cause the accuracy of ML-AQP to deteriorate as it is still able to generalize. Formally, we require workload/data updates to be significant so that the distributions at times $t$ and $t+1$ are no longer the same, meaning that the condition, $p_t(y)\neq p_{t+1}(y)$, holds for data updates and $p_t(\mathbf{m}) \neq p_{t+1}(\mathbf{m})$ for workload shifts. Although insertions/updates/deletions could be expected to be frequent, $p(y)$ might not change. %as these changes might happen in small regions in the data space affecting a small number of queries. 
Therefore, the key observation is that we do not need to track changes in the data space but instead need to monitor for changes in $p(y)$. To tackle both workload/data shifts we could naively retrain the models at fixed time intervals to be sure that the most updated queries and data are used. Over, 1M+ queries are executed daily in large deployments \cite{kandula2016quickraqp}; thus, it is easy/fast to find new queries executed on fresh data. However, we provide two alternative methods and in the next paragraphs we first address data updates and then workload shifts.

As shown previously, our solution is query-driven and does not access data at any time. Therefore, to detect changes in the  data/workload, we monitor queries that are successively executed by the data warehouse $(\mathbf{q}_t,\ldots)$. To detect changes to the aggregates distribution $p(y)$ we employ the two-sample Kolmogorov-Smirnov (KS) test. The KS test output statistic is $D = \sup_y |F_1(y)-F_2(y)|$, where  $F_1$ is the empirical CDF (ECDF) at time $t$ of answers $\mathcal{Y}_{1:t}$\footnote{The notation $1:t$ denotes the answers of queries at time-steps $1$ to $t$ and $t+1:$ denotes answers of queries from $t+1$ onwards.} of all queries that were used to train a model and $F_2$ is the ECDF of $\mathcal{Y}_{t+1:}$ from  monitored queries executed against the data warehouse. The KS test, evaluates the hypothesis that the two sets of answers come from the same distribution. The hypothesis is rejected at a significance level $\alpha$ if $D>c(a)\sqrt{\frac{n+m}{nm}}$, where $c(a)=\sqrt{-\frac{1}{2}+\ln{\frac{\alpha}{2}}}$ and $n=|\mathcal{Y}_{1:t}|$, $m=|\mathcal{Y}_{t+1:}|$ . If the hypothesis is rejected then the distribution has shifted and the model associated with the aggregate needs to be retrained. Retraining a model because of a distribution shift does not impose a large overhead as the cost of training the models is minimal,
(as shown in Section \ref{sec:training_overhead}). We are also considering other solutions such as data augmentation \cite{tanner1987calculation} to augment the existing set of queries and allow our models to generalize to possible data updates without retraining. 

We also need to address the case of a workload shift and employ a similar detection test. The workload is characterised by the query vectors $\mathbf{m}\in \mathbb{R}^{2d}$ and the KS test cannot be applied as their distribution $p(\mathbf{m})$ is multivariate. In this case we can apply Chebyshev's inequality which states that; given a random variable $X$ and its expected value $\mu$ then the probability of obtaining a sample point greater than $k$ standard deviations from the mean is $\leq \frac{1}{k^2}$, formally $Pr(|X-\mu|\geq k\sigma) \leq \frac{1}{k^2}$. 
This is clearly defined for a scalar value $X$ and we need the multivariate version of this as we have vectors $\mathbf{m}\in \mathbb{R}^{2d}$. For the multivariate case we can express the above as $Pr(\sqrt{(\mathbf{m}-\mathbf{\mu})^T\Sigma^{-1}(\mathbf{m}-\mathbf{\mu})}\ge k)\leq\frac{N}{k^2}$, where $\Sigma$ is the covariance matrix and $N=Tr[\Sigma^{-1}\Sigma]$. Hence, as queries are being monitored we can examine the inequality and trigger an alarm to retrain the models if the inequality is violated, meaning that if $Pr(\sqrt{(\mathbf{m}-\mathbf{\mu})^T\Sigma^{-1}(\mathbf{m}-\mathbf{\mu})}\ge k)\ge\frac{N}{k^2}$ then we need to retrain the models because of a workload shift.
\section{Evaluation}

\subsection{Experimental Setup}
\label{sec:experimental-setup}
\textbf{Datasets \& Workloads.}
For our experiments we used the following data sets and workloads:
\begin{enumerate}
    \item \texttt{TPC-H}\cite{tpch}: This is the standard TPC-H benchmark.
    \item \texttt{Instacart}\cite{instacart}: This is a data set of an online store. A database was created using the csv files which follows the setup of VerdictDB. \cite{park2018verdictdbaqp}.
    \item \texttt{Crimes}\cite{crimesdata}: This is a real data set of crimes reported in the city of Chicago. A workload for this data set was obtained from \cite{crimes_workload} which models a number of range-queries with multiple AFs. Their predicates are sampled from a number of random distributions to simulate various analysts executing queries at different subspaces of the data set. 
    \item \texttt{Sensors}\cite{sensors}: Is a data set comprised of a number of sensor readings including voltage, humidity temperature etc with a temporal dimension. A synthetic workload was created restricting the temporal dimension and extracting the \texttt{MIN(temperature)} and \texttt{MAX(humidity)}.
    % \item \texttt{synthetic}: A synthetic workload was constructed to stress test our chosen representation. The set of workloads include up to 100 attributes (stretching our meta-vector to 200, $\mathbf{m}\in \mathbb{R}^{200}$) with the number of set predicates up to $50$.
\end{enumerate}
\textbf{Training ML-AQP.} ML-AQP has a Training phase, much like sampling based AQP engines that need to create samples before being able to operate. 
For all workloads, we generate queries and train the models on $70\%$ of the complete workload. 
We then conduct the experiments and take measurements on the rest $30\%$. This is standard practise in the ML literature. 
For \texttt{Instacart}, we use a similar format of queries as VerdictDB \cite{verdict-instacart}. However, to facilitate learning we vary the predicate values. 
For all queries containing predicate values, 
we generate queries from the same template with values sampled from a normal distribution $\mathcal{N}(\mu, \sigma)$, 
where $(\mu, \sigma)$ is the average and standard deviation of the corresponding attribute, respectively. 
If the attribute contains a categorical value, 
the generated queries contain a value selected uniformly at random. The total number of training queries generated were $10^4$ and $3.3\cdot 10^3$ for testing. Some queries contained no predicates, 
in these cases no additional queries were generated.  The number of queries obtained is not large as typically there are millions of queries being executed on a daily basis in production environments \cite{kandula2016quickraqp}.

For \texttt{TPC-H}, we use a subset of the queries contained in the benchmark as we are still making progress on our \textit{Parser}. 
We generate $100$ queries for each of the queries used. 
A model is trained for each distinct AF. 
For \texttt{Instacart}, 
three different models are generated as three distinct AFs are used in this workload. For \texttt{TPC-H}, 
$12$ different models were generated. 
For \texttt{Crimes} and \texttt{Sensors}.
we generate a model per AF tested. 

\subsection{Performance}
We first examine the performance of ML-AQP and 
demonstrate the speedup gains over a popular 
database PostgreSQL. We compare the results using a sampling based AQP engine, VerdictDB \cite{park2018verdictdbaqp}. 
Let $t_b$ be the response time for PostgreSQL 
and $t_m$, $t_v$ the response times for ML-AQP 
and VerdictDB then speedup 
is measured by $\frac{t_b}{t_m}$ and $\frac{t_b}{t_v}$, respectively. 
For this experiment, 
we use \texttt{TPC-H} with 1GB and \texttt{Instacart} with its main fact tables (\texttt{order\_products}, \texttt{orders}) containing $~3$M and $~30$M rows. 
For \texttt{TPC-H}, 
we use a subset of all the template queries and for \texttt{Instacart}, 
we use the same format of queries as used in the evaluation of VerdictDB\cite{verdict-instacart}. 
For VerdictDB, 
uniform samples were created for large fact tables at 
$1\%$/$10\%$ ratio. This experiment ran on a single machine with an Intel(R) Core(TM) i7-6700 CPU @3.40GHz, 16GB RAM and 1TB HDD.

\begin{figure}[!htbp]
\begin{center}
\includegraphics[height=6cm,width=8.5cm, keepaspectratio]{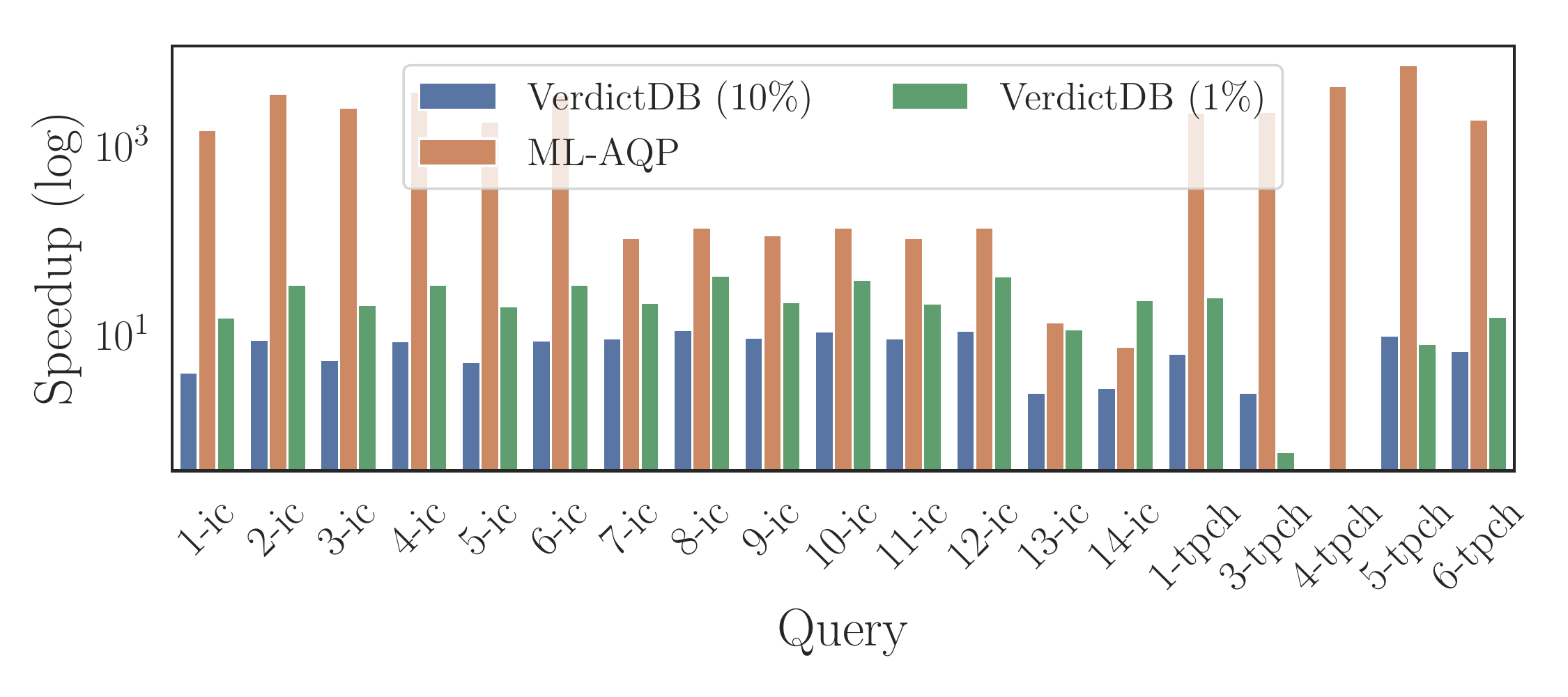}
\caption{Speedups offered by ML-AQP compared to VerdictDB}
\label{fig:speedup}
\end{center}
\end{figure}

The results are shown at Figure \ref{fig:speedup}. 
We can instantly notice that the speedup differences are huge (notice the log-scale on y-axis)\footnote{For $4$-tpch we could not get VerdictDB to execute this query.}. Even though we are using relatively small datasets, VerdictDB, understandably, cannot offer the same speedup as ML-AQP. The minimum/maximum speedup 
gained by ML-AQP is at $7\times$/$7200\times$,for VerdictDB $1\%$ $0.59\times$/$43\times$ (as we suspect that some of the computation is offloaded to the main engine) and for VerdictDB $10\%$ $3\times$/$11\times$. This stems from the fact that ML-AQP is only performing inferences at \textit{Prediction} mode using trained models. It does not need to scan any of the data at any time. To be more specific, Table \ref{tab:perfor-comparison} shows the mean response time along with the standard deviation and $95^\text{th}$ percentile for all queries across the four different systems. As it is evident, 
even at the $95^\text{th}$ percentile the response times for ML-AQP are no greater than $120$ milliseconds, satisfying the interactivity constraint set at $500$ms \cite{liu2014effects}.

\begin{table}[]
    \centering
    \begin{tabular}{l|lc}
    \toprule
    \textbf{System} & \textbf{Time} (sec) &   $95^\text{th}$ percentile (sec)   \\
    \midrule
    PostgreSQL & $1.62 \pm 1.21$ & $4.01$ \\ 
    VerdictDB (10\%) & $0.28 \pm 0.41$ & $0.79$ \\
    VerdictDB (1\%) & $0.14 \pm 0.3$ & $0.43$ \\
    ML-AQP  & $0.05 \pm 0.16$ & $0.12$ \\ 
    \bottomrule
    \end{tabular}
    \caption{Performance over all queries across systems}
    \label{tab:perfor-comparison}
\end{table}

Even for queries with relatively large \texttt{GROUP-BY}s the speedup is at $20x$. By default \texttt{GROUP-BYs} are a bottleneck in our case as multiple queries have to be executed for each distinct value of the attribute used in the \texttt{GROUP-BY} clause. For instance, query $14$-ic has approximately $50$K distinct values. In this experiment its values were cached as this would have been the default behavior. This is due to similar queries with the same GROUP-BY attributes being executed during \textit{Training} mode. 
% However, at Table \ref{tab:cached-distinctl} we present the overhead if ML-AQP had to execute a separate \texttt{DISTINCT} query to extract the values of the \texttt{GROUP-BY} attribute.
When caching the values, query $14-ic$ takes $0.67$ seconds to execute and $0.7$ seconds when it does not cache the GROUP-BY values. We can see minor impact, with an overhead attributed to the execution of the \texttt{SELECT DISTINCT} query at $0.3$ seconds. We can still get $7x$ better response times than PostgreSQL, where as VerdictDB ($1\%$/$10\%$) is at $24x/3x$ for this particular query. Although a larger speed-up is observed, for VerdictDB($1\%$) we will notice that accuracy using $1\%$ sampling ratio deteriorates with large errors in the individual groups.

\begin{figure}[!htbp]
\begin{center}
\includegraphics[height=3.5cm,width=5.5cm, keepaspectratio]{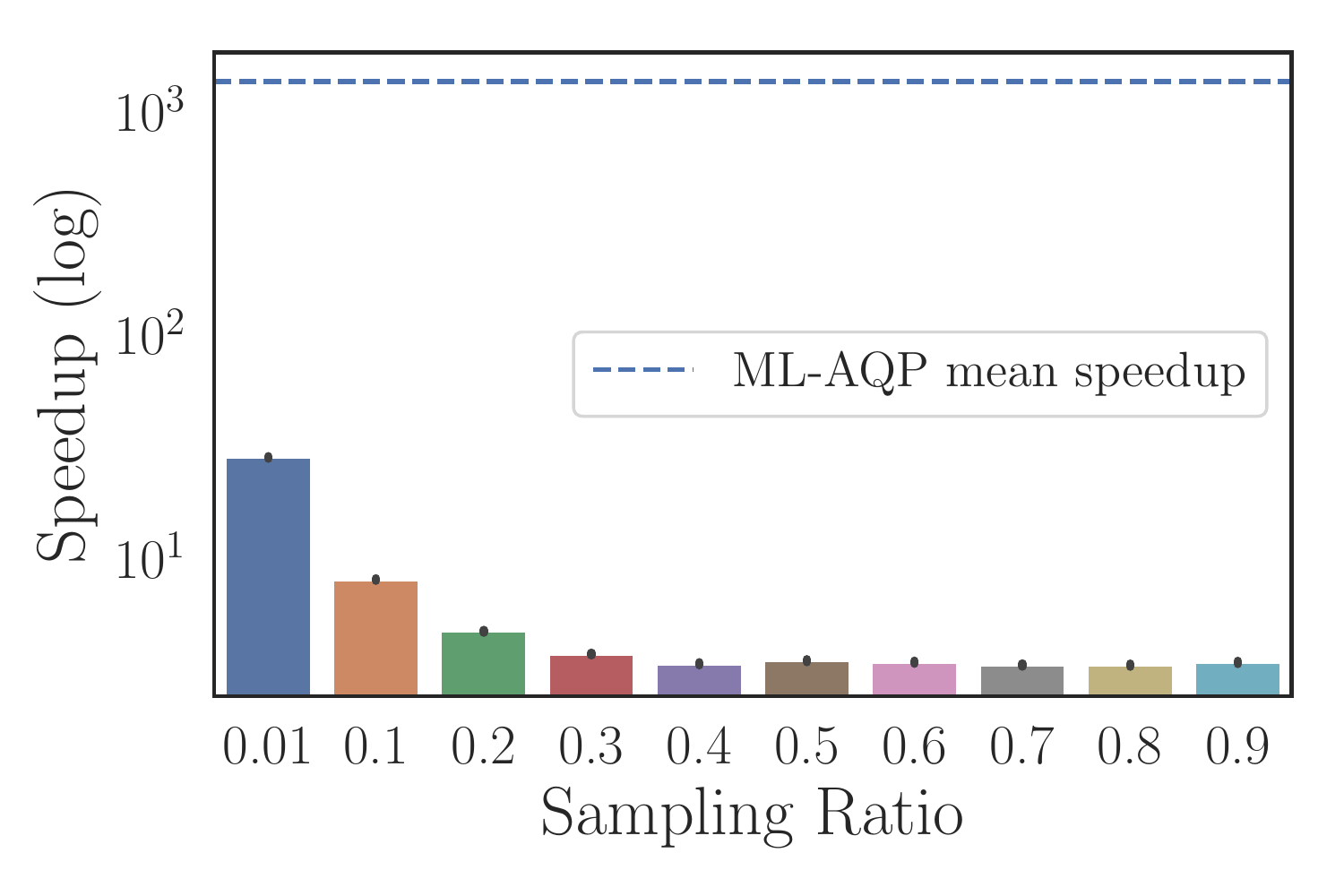}
\caption{VerdictDB speedup for an increasing sampling ratio.}
\label{fig:sratio-speedup}
\end{center}
\end{figure}

At Figure \ref{fig:sratio-speedup} we examine how the speed-up offered by sampling based and in general data-driven AQP solutions diminishes as the sampling ratio is increased. For point of reference we also provide the average speedup offered by ML-AQP which is not constrained by a sampling ratio as it uses no data. 

% \begin{table}[]
%     \centering
%     \begin{tabular}{cc|c}
%     \toprule
%         Cached \texttt{DISTINCT} & Non-Cached  & Speedup \\ 
%         & \texttt{DISTINCT} & (Non-Cached) \\ \midrule
%         0.67 (s) & 0.7 (s) & 7x (VerdictDB 3x)\\
%     \bottomrule
%     \end{tabular}
%     \caption{Overhead of calling \texttt{DISTINCT} over query $14$-iq}
%     \label{tab:cached-distinctl}
% \end{table}{}

\subsection{Performance at the Cloud}
Our solution is designed to alleviate the monetary, computational, storage costs in large deployments usually in the cloud. We first examine how 
the computational cost can be mediated using our solution. 
For this experiment, 
we use \textit{AWS Redshift}, 
with 2 \texttt{dc2.large} compute nodes 
and $1$ master node each at 16GB memory with 160GB SSD. 
We use a $100x$ scaled version of the \texttt{Instacart} dataset. The total storage footprint of this data set is $~100$GB with the main fact tables (\texttt{order\_products}, \texttt{orders}) containing $4.2$ billion and $0.5$ billion rows respectively. We execute the same \texttt{Instacart} queries \cite{verdict-instacart} and we set a \texttt{timeout} value at $5$(mins). After this, we abort the execution of the query. 
For VerdictDB we uniformly sample the same fact tables at $10\%$ ratio. In this experiment, 
we expect the results for VerdictDB to deteriorate. On the other hand ML-AQP is  constant in its performance as it is unaffected by data size. 
It is important to recall that the deployment of ML-AQP can happen in two ways: 
(i) All the models and required modules for ML-AQP can be distributed to all the analysts' machines and be loaded in memory during analysis (later experiments will showcase that the small storage footprint of ML-AQP permits this solution);
(ii) All models can be deployed at a server and be used as a service. This would have significantly lower costs than executing queries using Redshift. 
However, 
it might have more overhead as the predictions have to be transferred to the analysts machine over the network. 
We further examine the performance benefits of both solutions. 

\begin{figure}[!htbp]
\begin{center}
\includegraphics[height=6cm,width=8.5cm, keepaspectratio]{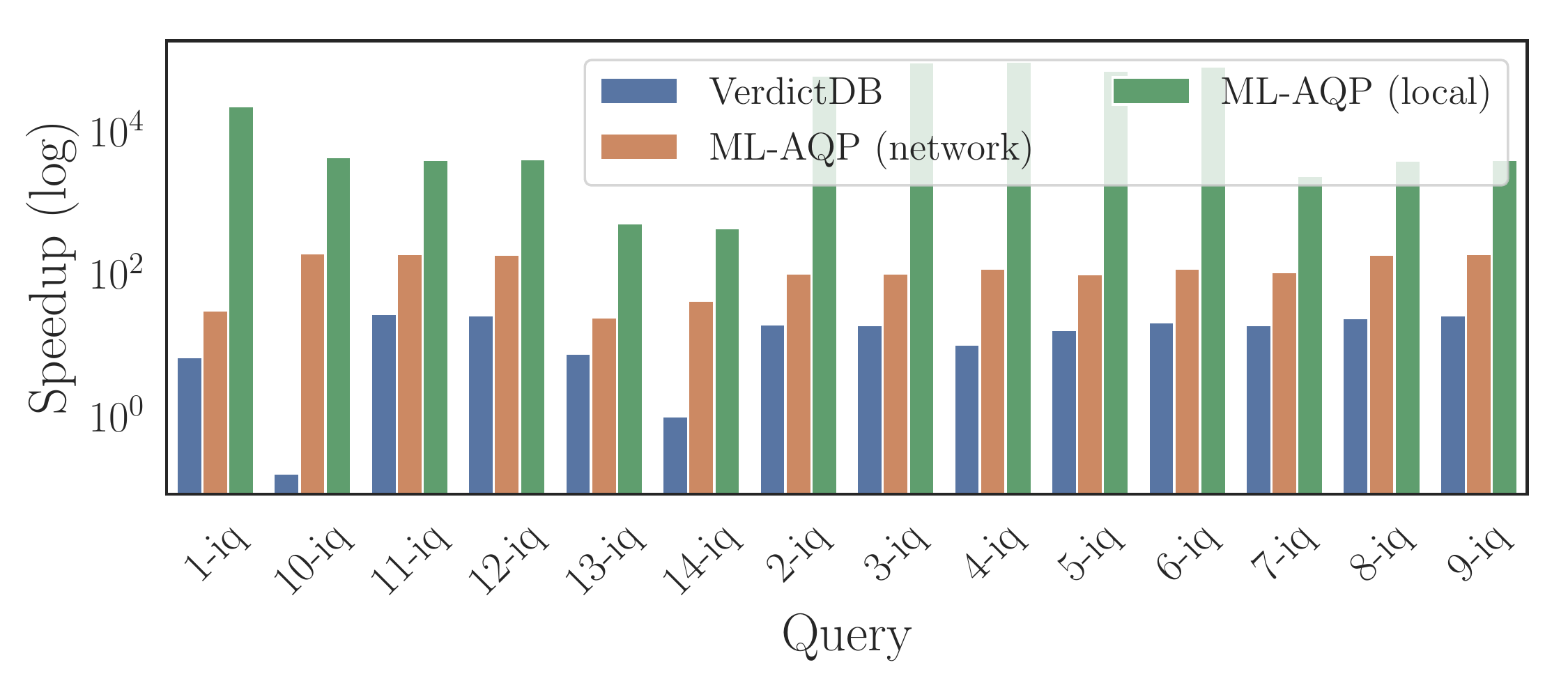}
\caption{Speedups in large deployments}
\label{fig:speedup-big}
\end{center}
\end{figure}

The results of this experiment are shown at Figure \ref{fig:speedup-big}. There are two different 
deployments for ML-AQP: 
(a) ML-AQP \textit{(network)}, 
(b) ML-AQP \textit{(local)}. 
For ML-AQP \textit{(network)} deployment, 
we set up a small server serving predictions over the network. It accepts HTTP POST requests with the extracted parameter values of the SQL query and returns a prediction of its answer. The results shown in Figure \ref{fig:speedup-big} are in \textit{log-scale}. As expected, 
the benefits of a \textit{local} deployment are far greater, although we would have to consider problems in maintaining the models as in this case ML-AQP are in each analysts machine. 
In addition, for some queries VerdictDB offers no speedups as Redshift is able to process those queries in an efficient manner. 

\begin{table}[]
    \centering
    \begin{tabular}{ccc}
    \toprule
       ML-AQP (\textit{local}) & ML-AQP (\textit{network}) & VerdictDB \\ 
        \midrule
        $443$-$10^5$ & $24$-$195$ & $0.16$-$28$\\
    \bottomrule
    \end{tabular}
    \caption{Minimum-Maximum speedups at the Cloud}
    \label{tab:minmax-speedups-big}
\end{table}{}
To be more concise, 
the min/max speedup benefits of the compared systems are shown at Table \ref{tab:minmax-speedups-big}. 
As evident, 
the \textit{local} deployment is orders of magnitude faster than both VerdictDB and an over the \textit{network} deployment. 
We also report on average response times and the response times at the $95^{\text{th}}$ percentile for all systems.
\begin{table}[]
    \centering
    \begin{tabular}{l|lc}
    \toprule
    \textbf{System} & \textbf{Time} (sec) &   $95^\text{th}$ percentile (sec)   \\
    \midrule
    Redshift & $55 \pm 75$ & $142$ \\ 
    VerdictDB  & $49 \pm 120$ & $300$ \\ 
    ML-AQP (\textit{network})  & $0.78 \pm 1.82$ & $2.68$ \\ 
    ML-AQP (\textit{local})  & $0.02 \pm 0.08$ & $0.25$ \\ 
    \bottomrule
    \end{tabular}
    \caption{Performance at the Cloud}
    \label{tab:perfor-comparison-big}
\end{table}
The results are shown in Table \ref{tab:perfor-comparison-big}. The first thing we notice, is that although VerdictDB has less mean response time than Redshift, at the $95^{\text{th}}$ percentile it is slower, possibly due to overheads of VerdictDB in deciding which samples to process. In addition, 
both the \textit{network} and \textit{local} deployments 
for ML-AQP offer mean sub-second latencies and only $2.68$ seconds at the $95^{\text{th}}$ for \textit{network}.

\subsection{Training Overhead}
\label{sec:training_overhead}
As stated earlier, 
ML-AQP has to go through \textit{Training} mode initially. At this stage, previously executed queries are used to train a variety of models and learn to predict the answers of future aggregate queries. Ideally training the models would happen \textit{locally} at Data Scientist's machines so as not to 
incur additional costs of repeatedly training and fine tuning the models in the Cloud. Therefore, in this experiment 
we measure the Training Time required to build a model 
with varying number of queries. 
We run this experiment locally on a single machine with an Intel(R) Core(TM) i7-6700 CPU @3.40GHz, 16GB RAM and 1TB HDD to demonstrate this capability. 
We compare this to the sample building time 
of VerdictDB with an increasing sample ratio. 
We use the \texttt{TPC-H} data set at $1$GB. At each iteration $3$ samples are built on the main fact tables. 
\begin{figure}[!htbp]
\begin{center}
\includegraphics[height=6cm,width=8.5cm, keepaspectratio]{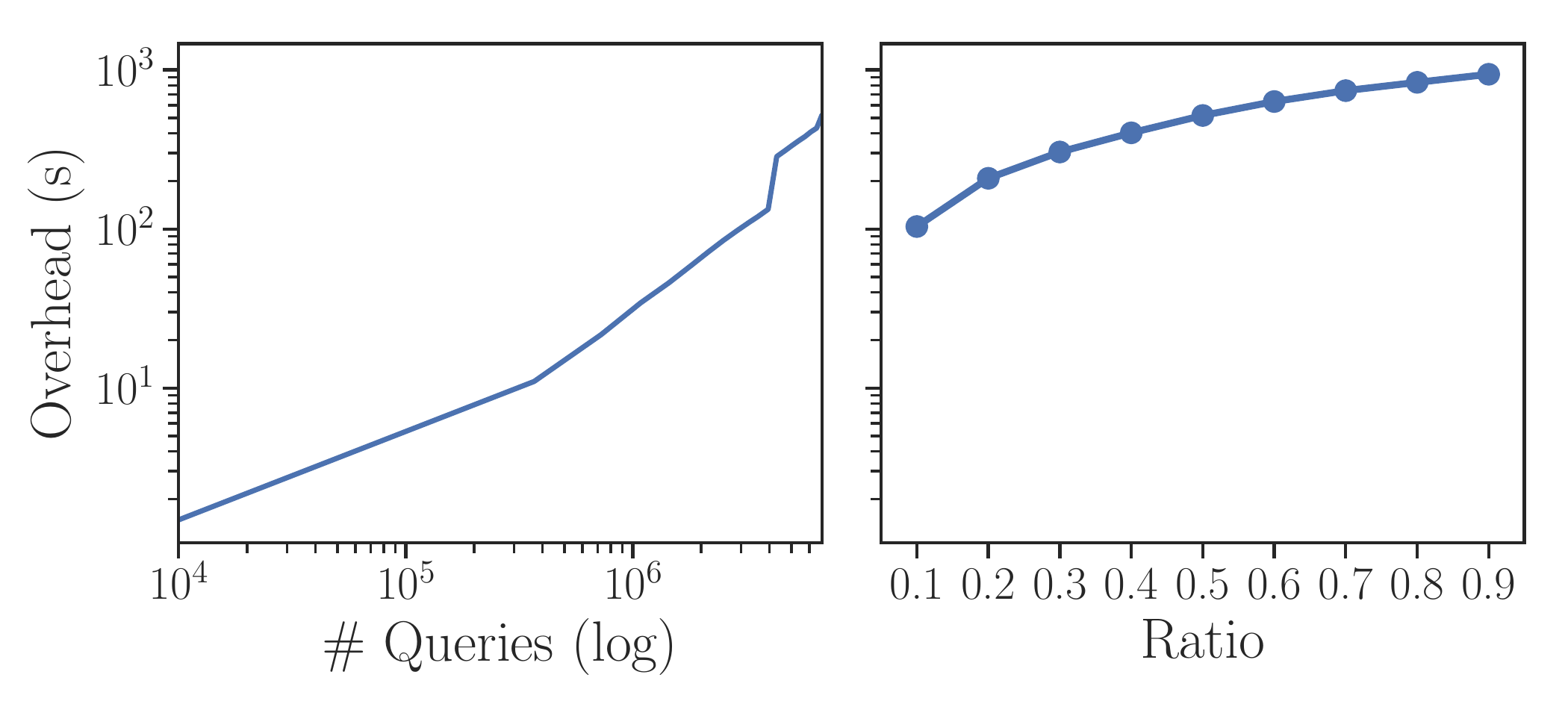}
\caption{(Left) Training overhead for an increasing number of queries (x-axis) (Right) Sample preparation time for VerdictDB}
\label{fig:training-overhead}
\end{center}
\end{figure}
Figure \ref{fig:training-overhead} shows the result of this experiment. The sample preparation time for VerdictDB, 
Figure \ref{fig:training-overhead}(right), increases linearly and even for $0.1$ ratio at $1$GB, takes longer than training ML-AQP on $4$ million queries. At $6+$ million queries, this overhead is still \textit{less} 
than the sample preparation time for VerdictDB at $0.6$ ratio. To put this in context, for ML-AQP to train on queries generated for \texttt{Instacart}, on $4,000$ for \texttt{AVG} queries took $1.1$ seconds and for \texttt{SUM} $3.3$ seconds. For \texttt{TPC-H} it took, less than a second to train each model. The only exception was for $\texttt{COUNT}$ as its associated queries included a large number of groups $(>50,000)$ and took $326$ seconds to train on around $5$ million training examples in total.
% Of course the time reported here is to train a single model. Ideally more models would have to be built. For instance, to account for every aggregate in the TPC-H workload $11$ models had to be built. For this, we could take advantage of training the models in parallel and not sequentially. 
% In addition, instead of training on the complete workload we could begin by experimenting on a subset of it. 
% We leave both of these methods as our future work. 
It is also important to note that sampling based AQP engines are susceptible to the size of the data set. In this experiment, 
we are only using $1$GB of data, as the size increases, 
sample preparation time is expected to increase, too. 
This would not be a problem for ML-AQP as it is only affected by the number of queries and it is \textit{not}, at any point, affected by the size of the underlying data set. In conclusion, both approaches, sampling-based AQPs and Query-Driven ML-based AQPs will have "\textit{training}" overheads. Their overheads are largely determined by different dimensions and as these solutions are designed to expedite query processing in \textit{petabyte} scale storage engines we expect ML-AQPs overhead to be much less.

\subsection{Accuracy}
\subsubsection{Accuracy on TPC-H \& Instacart}
To assess the accuracy of ML-AQP we measure the \textit{Relative Error}\cite{park2018verdictdbaqp,park2017databaseaqp,agarwal2013blinkdbaqp,kandula2016quickraqp} across all the query templates of both \texttt{Instacart} and \texttt{TPC-H}. ML-AQP was trained on past queries generated as described in Section \ref{sec:experimental-setup}. Three models  were trained using \texttt{LightGBM} to answer \texttt{Instacart} queries, one for each AF involved. For \texttt{TPC-H}, 11 models were trained using \texttt{XGBoost} as the queries were largely referring to AFs on different attributes. The number of rounds were set to $10^4$ with early stopping when no more improvement was shown. Objective was set to \texttt{squared\_error}. We compare our results with VerdictDB, which created samples over the large fact tables at ratios of $1\%$/$10\%$. 
\begin{figure}[!htbp]
\begin{center}
\includegraphics[height=6cm,width=8.5cm, keepaspectratio]{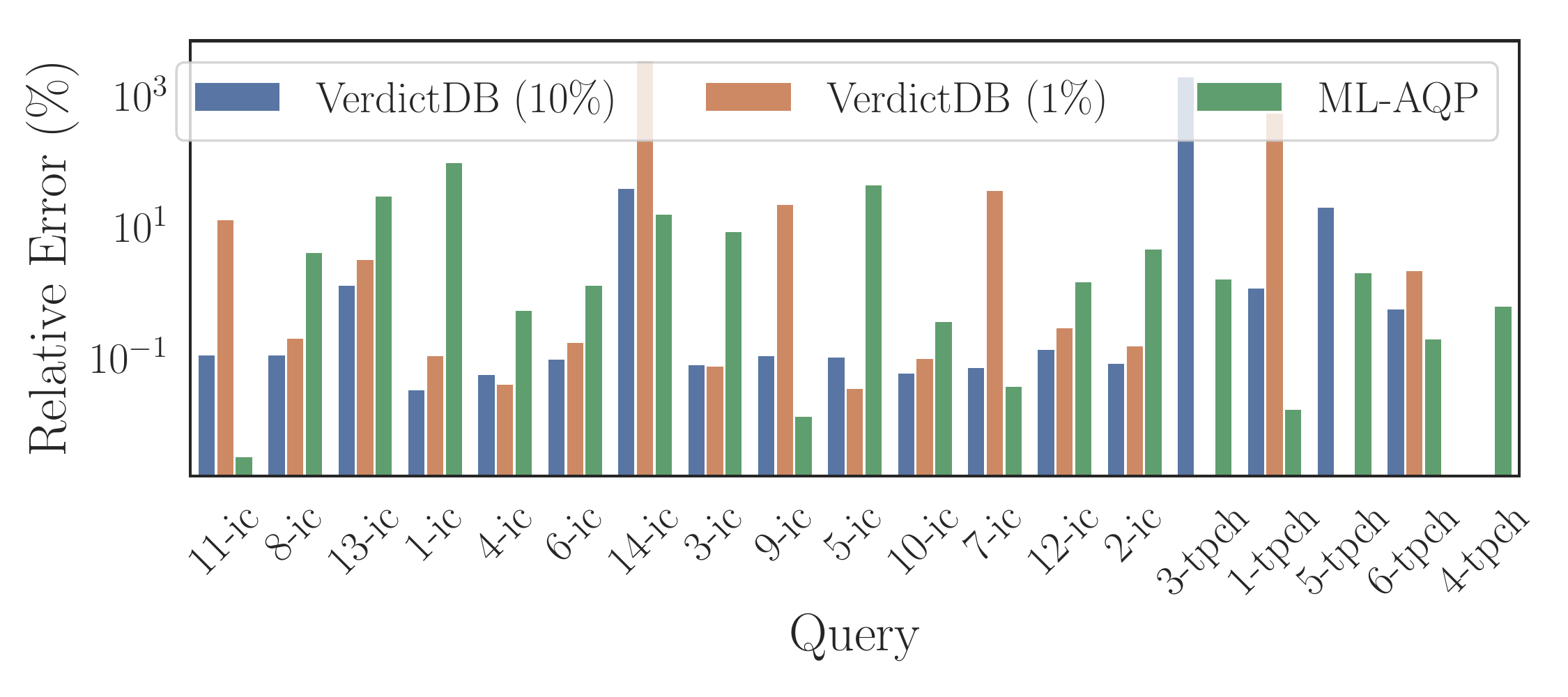}
\caption{Relative Errors for each query in \texttt{TPC-H} \& \texttt{Instacart}}
\label{fig:rel-error-each}
\end{center}
\end{figure}
An initial set of results is shown at Figure \ref{fig:rel-error-each}. A first impression is that both systems perform really well over a large range of queries. Please note that the results show the average relative error over each query template. Where for each query template multiple, $>50$, queries were executed with random predicate values as described in (\ref{sec:experimental-setup}). ML-AQP is able to accurately answer $80\%$ of queries for \texttt{Instacart} and $100\%$ of the selected \texttt{TPC-H} queries with relative error below $10\%$. We can also visually discern that ML-AQP outperforms VerdictDB for many queries. VerdictDB at $1\%$ was not able to answer accurately queries that have a large number of groups, such as $14$-ic and in some cases the groups returned by VerdictDB did not match the ones returned by the engine ($3$-tpch, $5$-tpch)\footnote{We were not able to run query $4$-tpch as an uknown error was thrown at runtime.}.  For ML-AQP, queries \textit{1-ic} and \textit{5-ic} produce large relative errors. This is understandable as these queries include no predicates and are simply the results over a full scan of the table (ie are simple \texttt{SELECT AF FROM T}). The results of such queries can easily be cached. ML-AQP is not expected to answer such queries as the meta-vector $\mathbf{m}$ is filled with \texttt{nan} values, to which the model simply ignores as there are no patterns to be learned.  

\begin{figure}[!htb]
\begin{center}
\begin{tabular}{cc}
\includegraphics[height=4cm,width=4cm,keepaspectratio]{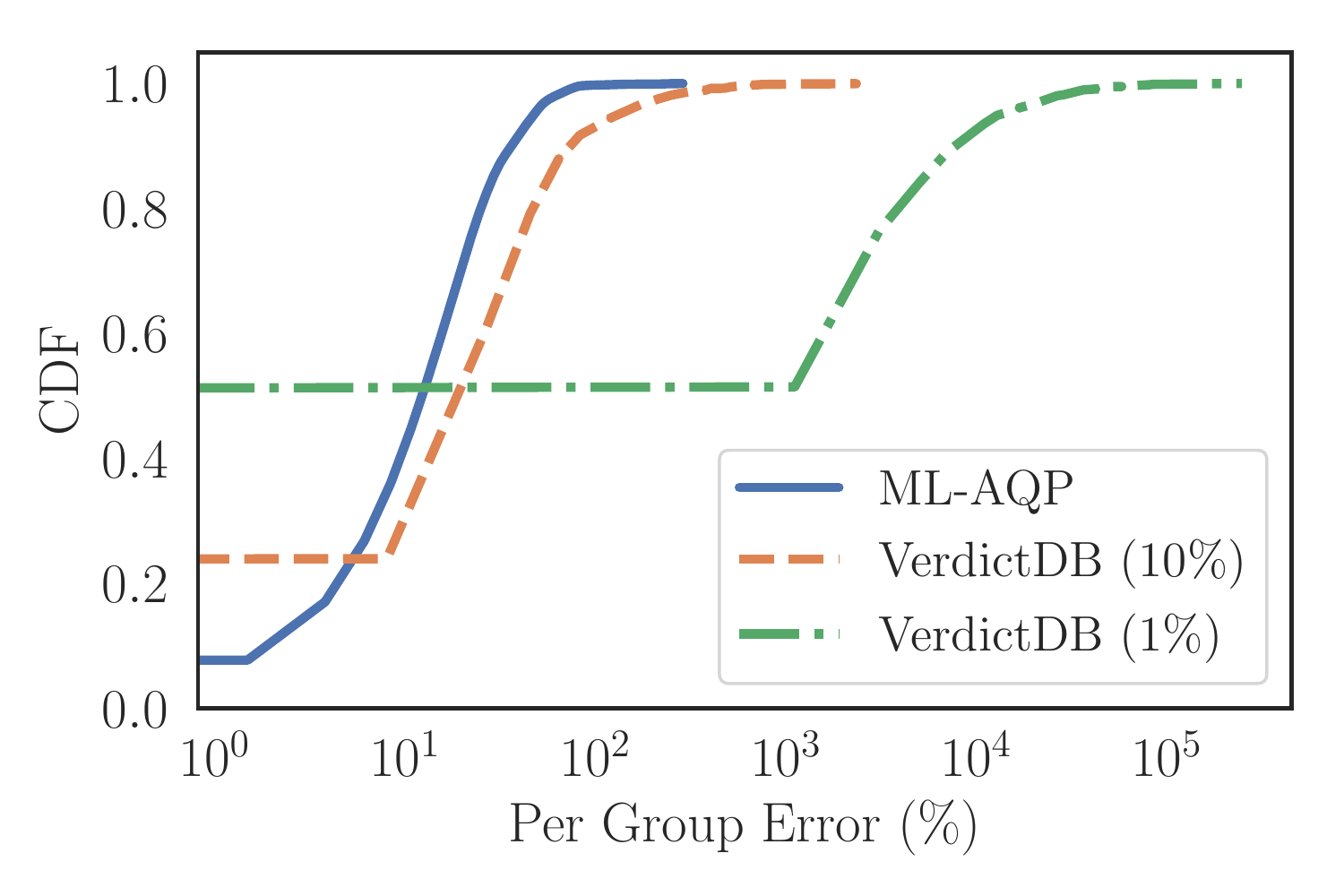}
\includegraphics[height=4cm,width=4cm,keepaspectratio]{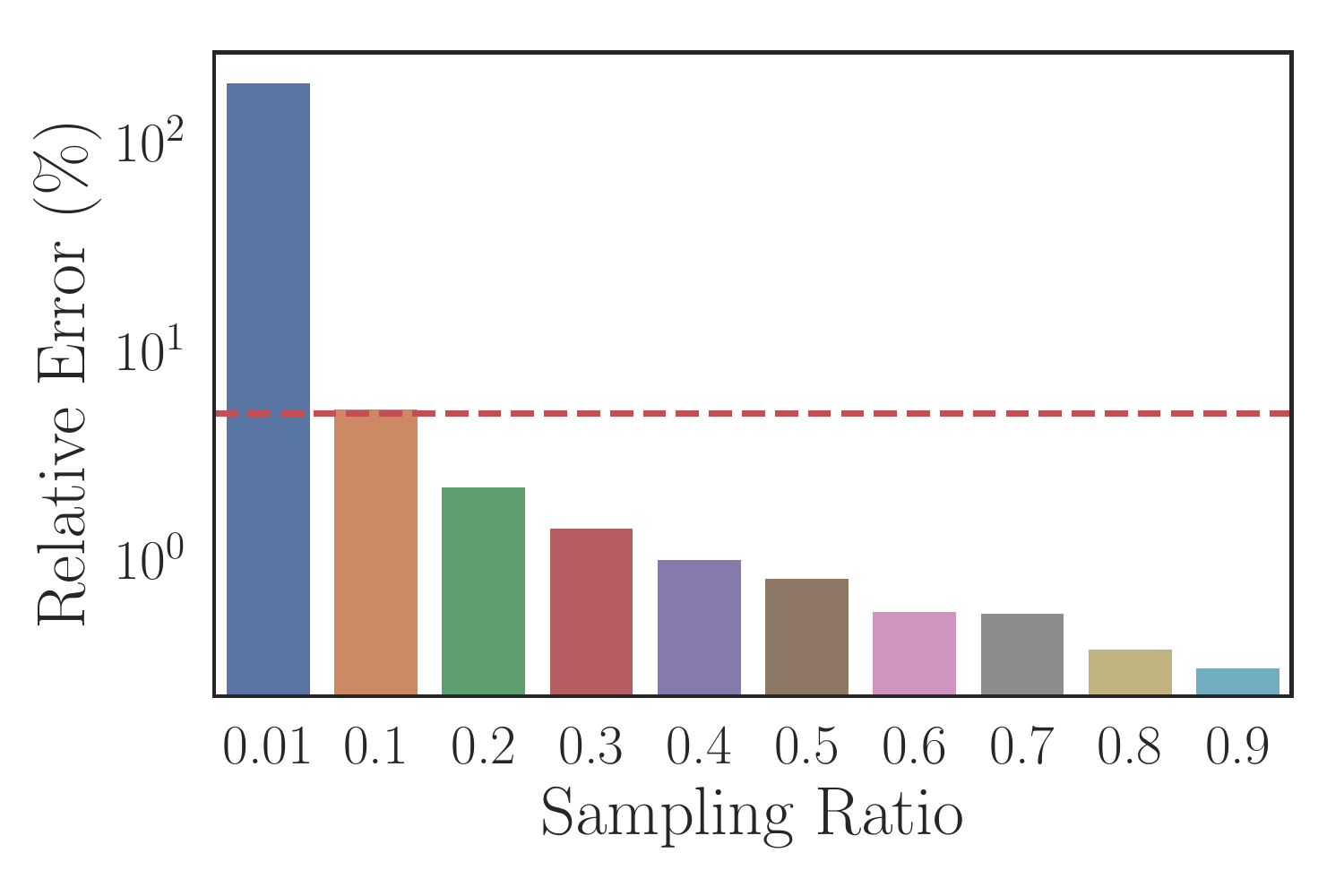}\\
\end{tabular}
\caption{(Left) CDFs of relative error per group in a \texttt{GROUP-BY} query (right) Relative error for an increasing sampling ratio with the mean relative error for ML-AQP as a horizontal line.}
\label{fig:accuracy-cdfs}
\end{center}
\end{figure}

We also measure the relative error on a per-group basis. Especially, for query $14$-ic where a large number of groups are present we notice high relative error. Figure \ref{fig:accuracy-cdfs}(left), shows the CDF of the relative error across groups for all three systems. ML-AQP outperforms VerdictDB at both sampling ratios ($1\%$, $10\%$), which shows that it can accurately estimate the aggregates across groups. Given the prior discussion we do not mean to say that sampling based engines have less accuracy. Instead, at a small sampling ratio the benefits are not great and the large trade-off between accuracy and speed makes their use inappropriate. Hence, sampling based engines can be used in parallel to ML-AQP, when the analyst needs more accurate answers and they are willing to sacrifice some of the efficiency for it, as also suggested by Figure \ref{fig:system-placement}. So, the systems can co-exist if we use sampling based engines with a higher sampling ratio as the expected error over all queries decreases as is shown at Figure \ref{fig:accuracy-cdfs}(right).

\subsubsection{Accuracy on range queries over spatio-temporal data.}
\begin{figure}[!htbp]
\begin{center}
\includegraphics[height=6cm,width=8.5cm, keepaspectratio]{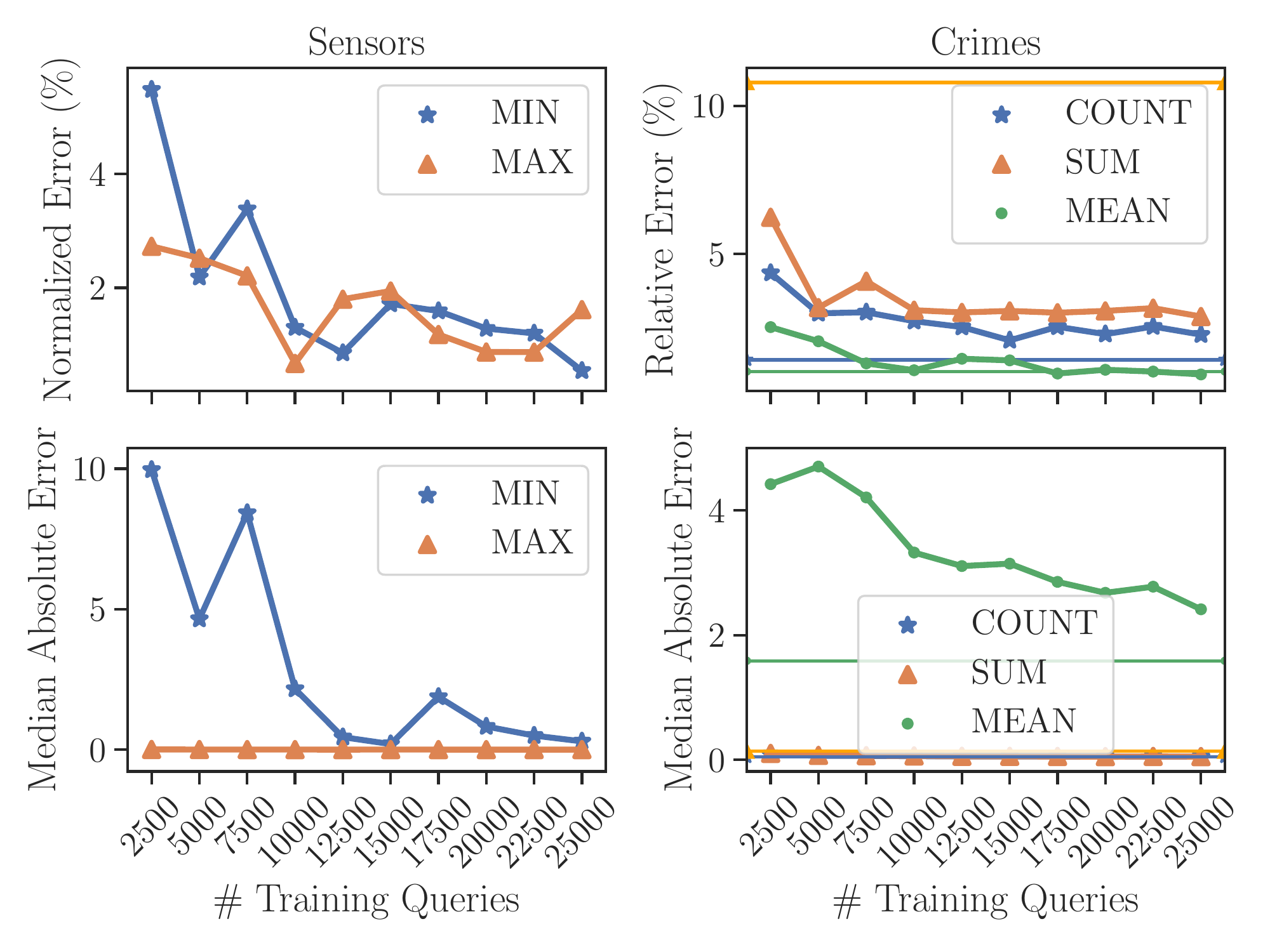}
\caption{Accuracy over \texttt{Sensors} and \texttt{Crimes} for an increasing number of \textit{training} queries and over different AFs. (Top) Relative/Normalized Error (Bottom) Median Absolute Error. (Right Column) For Crimes the accuracy of VerdictDB is plotted as horizontal lines.}
\label{fig:accuracy-crimes-sensors}
\end{center}
\end{figure}
We have also measured the accuracy of ML-AQP in predicting the responses of range-queries over spatio-temporal data sets. Specifically, multiple synthetic queries are executed over \texttt{Crimes} restricting its spatial dimensions and returning a response \texttt{COUNT}, \texttt{MEAN} or \texttt{SUM} over other attributes included in the data set. 
Namely \texttt{COUNT} returns the number of recorded incidents within the defined area, \texttt{MEAN} 
is the average \textit{Beat} number which is a police defined number describing the area, and \texttt{SUM} of the arrests over the specified area. 
For \texttt{Sensors} we restricted the temporal dimensions and extract the \texttt{MIN}(temperature) and \texttt{MAX}(humidity). All of the results at Figure \ref{fig:accuracy-crimes-sensors}(top) show that the relative error is below the targeted $10\%$ for this kind of data sets and conntinues dropping as more queries are being used for training the models. We also plot the accuracy for VerdictDB ($1\%$) as horizontal lines only for the \texttt{Crimes} data set as VerdictDB does not support \texttt{MIN}/\texttt{MAX} aggregates. 
Note that for \texttt{Sensors}, 
we report on the Normalized Error $|\frac{y-\hat{y}}{\overline{y}}|$, which computes the absolute difference divided by the mean response. 
The reason is that for this workload, 
the values are really small ($\mu=9$) and the measured relative error is not robust as it might report a $50\%$ error even if $y=2$ and $\hat{y}=1$. This is also encountered in \cite{kandula2019experiencesaqp} and similar technique is employed. To provide more context as to how close the predictions are in relation to the true response, we also provide results on the Median Absolute Error (MAE) at Figure \ref{fig:accuracy-crimes-sensors}(bottom). It is a well known metric in the ML community that is robust to outliers indicating the median of the absolute error between $y$ and $\hat{y}$. As evidenced, 
the absolute difference is small for all 
aggregates and data sets and continues to drop 
as more queries are used for training. The accuracy obtained is similar to VerdictDB's with ML-AQP having lower relative error for \texttt{COUNT}. In addition, we stress the fact that we are able to predict the responses for \texttt{MIN} and \texttt{MAX} that to our 
knowledge are not supported by most AQP systems. 
In addition, as the number of queries increase, 
we see a drop in relative error suggesting that more accurate predictions can be obtained. 
Overall, the results of this experiment show that ML-AQP is able to support a wide variety of aggregates over a diverse set of data sets. 

\subsubsection{Accuracy on error estimation}
We study the effectiveness of the prediction intervals constructed using QR. For this experiment we train two \texttt{LightGBM} models on \texttt{quantile loss}, with parameters \texttt{n\_estimators}$=1500$ and \texttt{l\_rate}$=0.001$. We set $t=0.95$ and train the two models using \texttt{alpha}$=1-t$ and \texttt{alpha}$=t$. This effectively creates a prediction interval that would ideally provide a coverage rate of $90\%$. Coverage rate is used in other work to assess prediction intervals \cite{lei2018distributionconformal,foygel2019predictive} and is essentially an empirical estimate of the predictions that will fall within the proposed interval. It is computed using a held-out set of queries. Specifically, the two models generate responses for $\hat{y}_{5^\text{th}}$ and $\hat{y}_{95^\text{th}}$. We test each \textit{true} value $y$ on $\hat{y}_{5^\text{th}}\leq y \leq \hat{y}_{95^\text{th}}$ and report the ratio of queries where the condition is true. We used the queries of \texttt{Instacart} and conduct this experiment on three different AFs \texttt{COUNT}, \texttt{SUM}, \texttt{AVG}.

\begin{figure}[!htbp]
\begin{center}
\includegraphics[height=3.5cm,width=5.5cm, keepaspectratio]{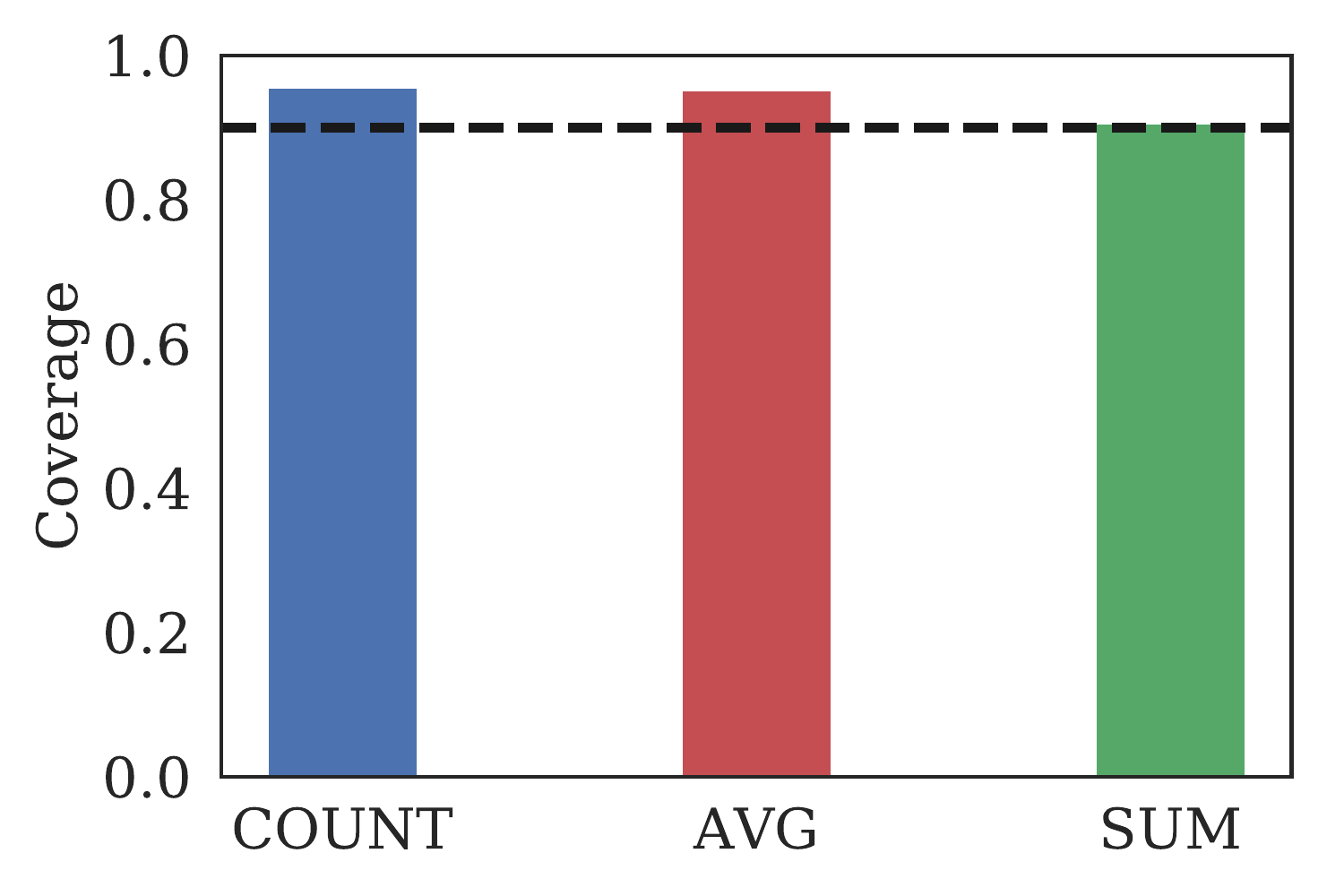}
\caption{Coverage Ratio for different AFs. Horizontal line drawn at $90\%$}
\label{fig:error-estimation}
\end{center}
\end{figure}

The results at Figure \ref{fig:error-estimation} confirm, that empirically a value lies within the interval, provided by QR estimates, by an estimated probability $>0.9$. In short, using QR, ML-AQP is able to provide good probabilistic intervals for the true answer. Using this interval the user can choose whether they trust the prediction or they wish to get a more accurate estimate using an S-AQP or the data warehouse engine. 

\subsection{Storage}
For this experiment, 
we measure the \textit{Storage} overhead of ML-AQP. 
At the end of the Training phase we deploy ML models at analysts devices or a central device. 
Measuring the storage overhead and ensuring that 
this is adequately small is of great importance. 
We expect orders of magnitude smaller storage footprint than sampling based AQP engines as we neither store any of the data nor any of the queries used for training. 
We initially examine how much memory is required by a model with increased complexity. The main factor contributing to the size of the selected ML models (GBMs) is the number of trees and their depth. 

\begin{figure}[!htbp]
\begin{center}
\includegraphics[height=3.5cm,width=5.5cm, keepaspectratio]{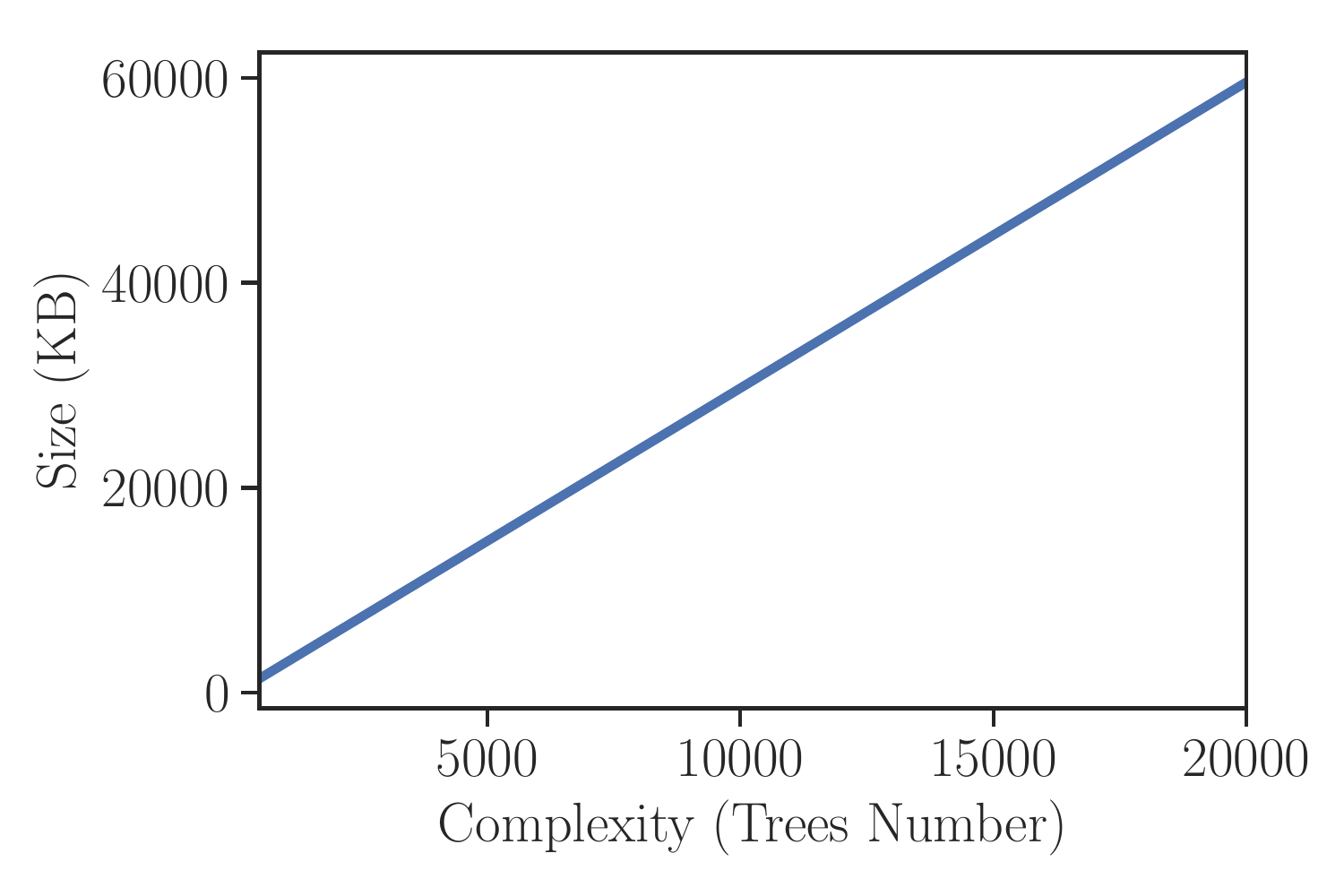}
\caption{Increasing number of trees (x-axis) and size in kilobytes}
\label{fig:size-models}
\end{center}
\end{figure}

Figure \ref{fig:size-models} shows an increase in the total storage required by an increasing number of trees. For conducting the experiments over \texttt{Instacart} and \texttt{TPC-H} the number of trees never exceeded $1700$ with some AFs requiring as little as $100$ trees. ML-AQP requires additional 
storage for encoding categorical values and for caching values obtained from queries with \texttt{GROUP-BY} clauses. 
For instance for \texttt{Instacart}, there are $50,000$ categorical values and ML-AQP requires an extra $4$MB (on top of the storage required by the models). This cost increases linearly as the number of labels increase. 
Accounting for all of this and even any required modules by the implementation of ML-AQP still does not match the storage overhead required by sampling based AQP. 
To put this in context, \texttt{Instacart} requires $2.4$GB of storage for its tables. To sample its main fact tables \texttt{orders} and \texttt{order\_products} at $0.1$, VerdictDB required $1.8$GB in total. 
On the other hand, ML-AQP requires a mere $15.5$MB to cover the aggregate queries issued against \texttt{Instacart}, this includes all models and catalogues. Given this information, 
we can safely assert that ML-AQP is extremely light-weight and can easily reside in main memory during analysis. 
% Note: we would like to draw the attention of the 
% reader to recent methods for compressing ML models achieving the same accuracy, like knowledge distillation \cite{knowdistil}; 
% this is expected to result to significantly smaller ML models (in size) shaping our future research agenda in this direction of research.

\subsection{Updates Adaptation}
\begin{figure}[!htbp]
\begin{center}
\includegraphics[height=3.5cm,width=7cm, keepaspectratio]{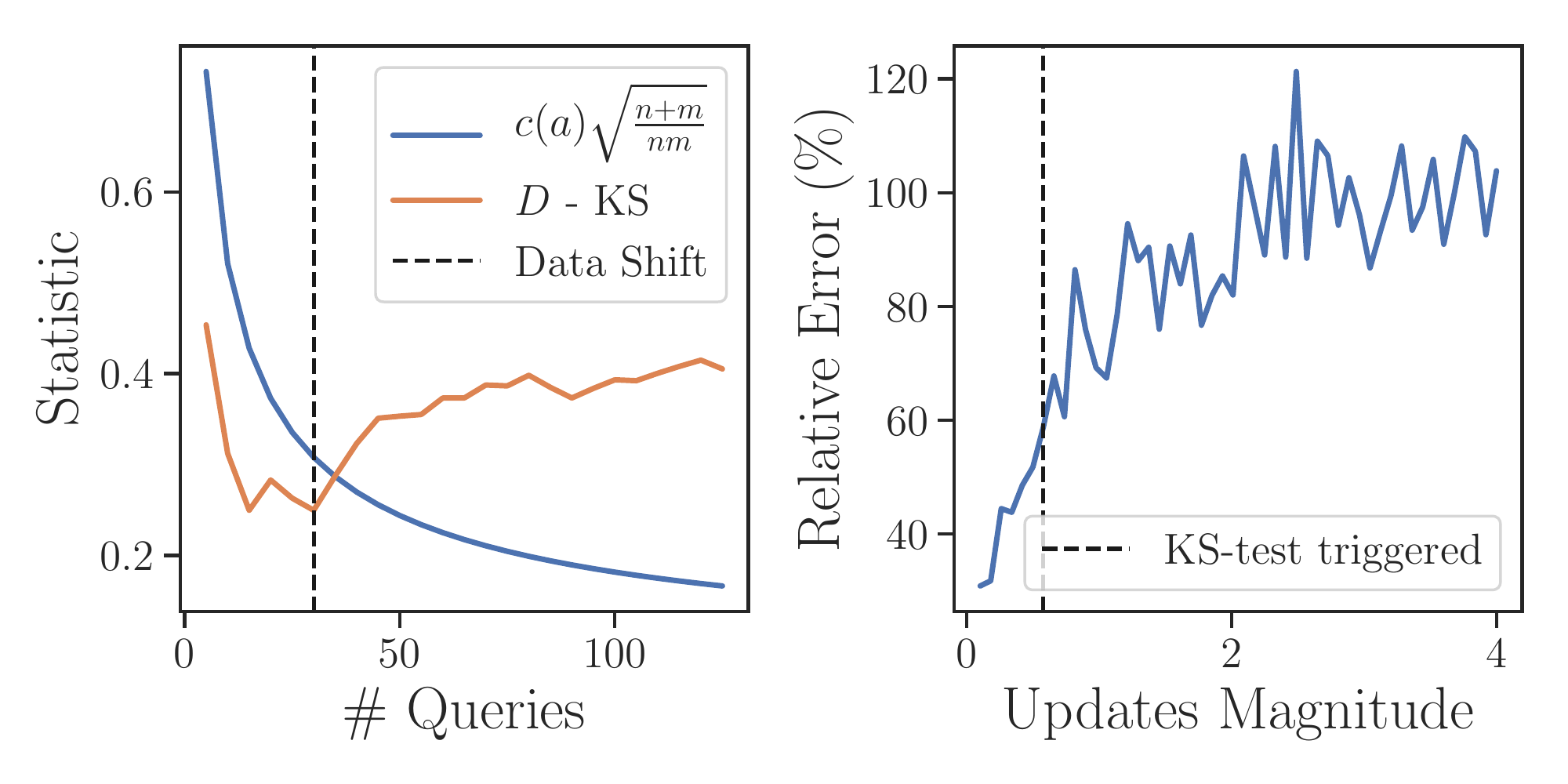}
\caption{(Left) x-axis; Number of queries processed initially from $p_1$ and then from $p_2$; y-axis; measured statistic (Right) Updates Magnitude and firing of the KS statistic before relative error increases.}
\label{fig:data-shift}
\end{center}
\end{figure}

We conduct an experiment to assess ML-AQP's detection mechanisms for updates in data and workload. As we have elaborated we need to identify cases where $p(y)$ and $p(\mathbf{m})$ change by observing queries executed at the data warehouse without accessing any data. Monitoring actual insertions/updates/deletions could prove futile as the distribution of $y$ might not be changing. For this experiment we use two distributions $p_1(\mathbf{m},y)$ and $p_2(\mathbf{m},y)$. The distributions for parameters $\mathbf{m}$ are multivariate Normal distributions initialized randomly at the data space of \texttt{Crimes} data set and the answers $y$ are the actual answers for \texttt{COUNT} over the same data set. We first test for data shift detection. In this experiment we obtain the empirical distribution function of $p_1(y)$ and conduct the KS test at regular intervals. At a specific point in time we change to $p_2(y)$ which we then expect that KS statistic will go over the threshold. The results at Figure \ref{fig:data-shift}(left) show that as the distribution shifts the KS statistic increases and becomes larger than $c(a)\sqrt{\frac{n+m}{nm}}$ as soon as the data shift happens (vertical dotted line). In addition, Figure \ref{fig:data-shift}(right) shows when the KS statistic fires relative to the impact of updates on $p(y)$ indicating that the model would be updated before relative error increases dramatically.

\begin{figure}[!htbp]
\begin{center}
\includegraphics[height=4cm,width=5.5cm, keepaspectratio]{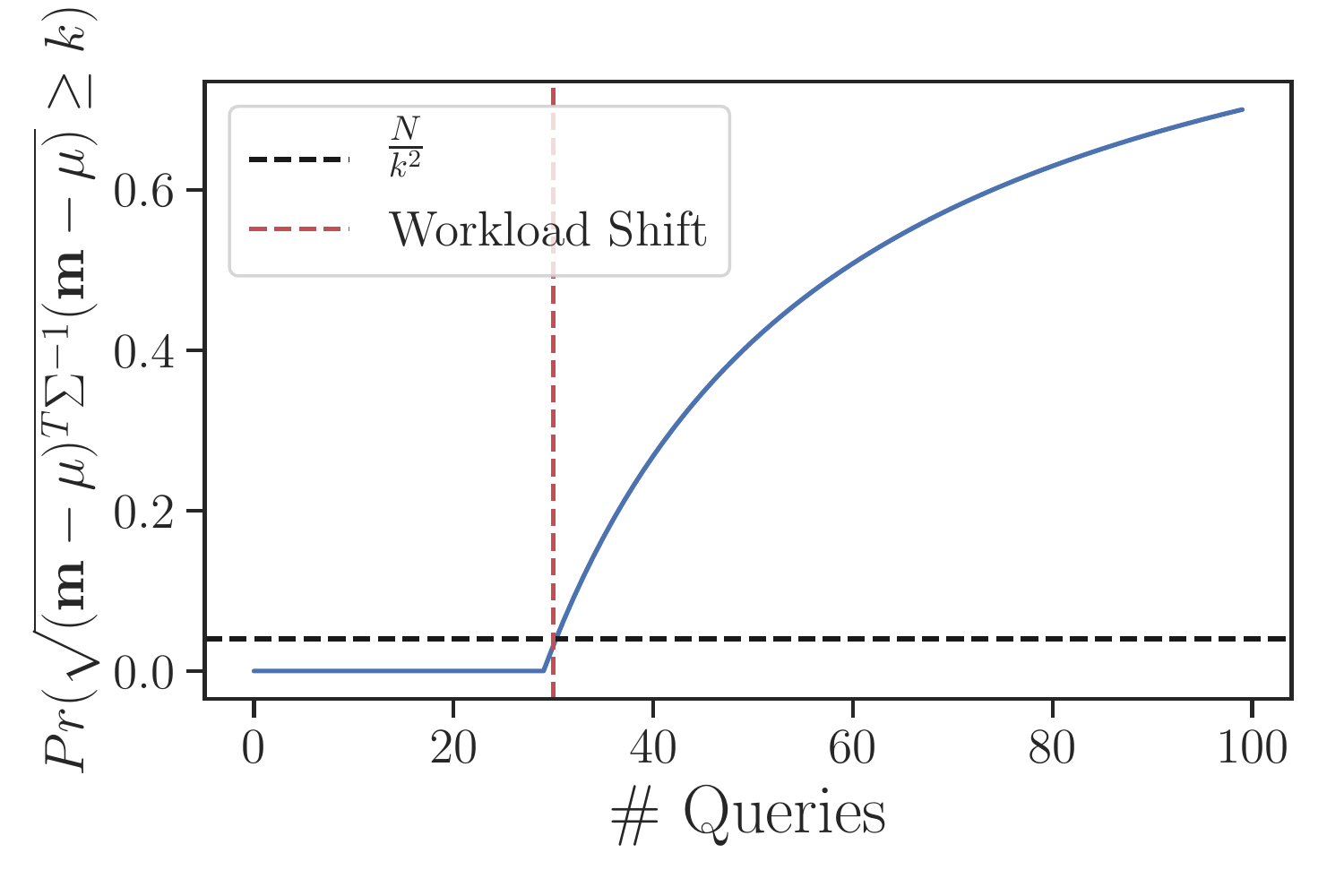}
\caption{Number of monitored queries and  probability of exceeding $k$ standard deviations along with Chebyshev's inequality threshold.}
\label{fig:workload-shift}
\end{center}
\end{figure}

We perform a similar experiment to assess the detection of a workload shift. Again,we have two different distributions, where initially queries are observed from the first distribution and at a specific point in time queries are sampled from the other distribution. We monitor Cheybyshev's inequality and detect whether the inequality has been violated to trigger a workload shift. The results shown at Figure \ref{fig:workload-shift} denote that as the workload changes (indicated by the vertical dotted line) the threshold (horizontal dotted line) provided by the inequality is exceeded and we can successfully detect whether the patterns have shifted and adapt accordingly. In general, in the face of data or workload updates ML-AQP is able to appropriately detect and adapt to such changes.

\section{Related Work}
%normal data analytic engines
To meet the needs of interactive query processing in large analytic environments various big data engines \cite{armbrust2015spark,dean2008mapreduce,zaharia2016apache} and columnar databases \cite{gupta2015amazon} have been developed. However, the goal of truly interactive analysis still remains elusive, as such engines produce \textit{exact} results. Their results have to be computed over large quantities of data. Therefore, research in AQP has been pretty strong the last decades \cite{zeng2015gaqp,potti2015daqaqp,olma2019tasteraqp,acharya1999aquaaqp,park2017databaseaqp,park2018verdictdbaqp,agarwal2013blinkdbaqp,kandula2016quickraqp,kandula2019experiencesaqp,hellerstein1997onlineaqp,chaudhuri2017approximateaqp, babcock2003dynamicaqp,ma2019dbest, garofalakis2001approximate} and still the list is not exhaustive. 
%aqp engines sampling based and online based
We can categorize most AQP engines to sampling based AQP engines \cite{agarwal2013blinkdbaqp,park2018verdictdbaqp,olma2019tasteraqp} and online aggregation engines \cite{zeng2015gaqp, hellerstein1997onlineaqp}. Sampling based AQP engines create samples over some (or all) of the columns in tables and produce answers with error guarantees based on samples. On the other hand, online aggregation engines produce a result as quickly as possible and then keep on refining it as more and more data are processed until the user stops the query execution. However, AQP is struggling to find its way through the industry. A notable approach is QuickR developed from Microsoft \cite{kandula2016quickraqp, kandula2019experiencesaqp} but there is more work to be done.

%ML models in data management and ML models in AQP
More recently we see the deployment of ML models to a wide variety of data management problems: selectivity estimation \cite{dutt2019selectivity,cardinalitydeeplearning,anagnostopoulos2017query,anagnostopoulos2015learning,ortiz2019empirical,kipf2018learned,sun2019end,yang2019selectivity,woltmann2019cardinality}, query optimization \cite{neo}, in which ML is used to decide on a query plan, or to create indexes \cite{kraska2018case}, for expediting visualisation \cite{wang2018neuralcubes} and to AQP \cite{ma2019dbest,hilprecht2019deepdb, thirumuruganathan2019approximate, kulessa2018model}. We believe that this approach can be fruitful if used with care. This is why we are not aiming to replace already existing AQP engines or data analytic systems and instead provide an addition to the stack. We believe the user needs to have a choice considering the trade-offs between speed and accuracy. As such our approach is similar to the recent trend of applying ML over data management problems, in that we employ ML models for AQP. Approaches such as \cite{ortiz2019empirical,cardinalitydeeplearning,kipf2018learned,sun2019end,yang2019selectivity,woltmann2019cardinality} make use of ML to estimate cardinalities for their use in a query optimization setting. Similar methodologies are adopted in that first features are extracted either from data or from queries and subsequently used for training models. This is to be expected as any relevant work that makes use of ML needs to follow this process. However, all of the aforementioned works focus on cardinality estimation and not on AQP. In addition they follow different modelling/vectorization for the queries, use different ML models and none of the works explicitly address data/workload updates and error guarantees. 
Perhaps the most similar work to ours is NeuralCubes (NC) \cite{wang2018neuralcubes} in which the authors describe a query-driven ML-based system to be used as a visualisation backend engine using AQP (COUNT/AVERAGE) answers. Compared to NC, ML-AQP (i) provides error guarantees, (ii) support updates in both data and workload, (iii) support more AFs, (iv) explicitly provide information on training/storage/prediction-serving overheads, and (v) provide all that with inherently simple, easy to understand ML models. We have conducted experiments based on two data sets that were used in NC and report on Relative Absolute Error (RAE) which is the metric used in NC. ML-AQP's RAE for \texttt{BK-Austin} was $3.9\%$ (where NC is at $3.88\%$) and for \texttt{BK-NYC}, RAE was $3.6\%$ (where NC is at $4.25\%$). The results suggest better or similar accuracy over NC with a much smaller training overhead--ML-AQP trains its models $>10X$ faster than NC on weaker hardware.

Compared to other approaches focusing on ML for AQP such as  \cite{ma2019dbest,hilprecht2019deepdb,thirumuruganathan2019approximate,kulessa2018model} ML-AQP neither learns from data nor uses data to construct samples or models. ML-AQP employs a novel query-driven method, based on vectorized representations of previously executed queries and their results and is oblivious to the underlying data distribution. In addition, ML-AQP's focus is not solely in COUNT/SUM/AVG as most works \cite{hilprecht2019deepdb,thirumuruganathan2019approximate,kulessa2018model} but offers support for any kind of AF through its AF agnostic methodology. Nevertheless, data-driven approaches are surely a promising avenue and we believe all of these approaches could complement each other. In cases where;i) data sets are massive, ii) no models or samples can be built and stored efficiently, and iii) there is a low cost requirement, ML-AQP appears to be more favorable.

\section{Conclusions}
In this paper we described ML-AQP, %a query-driven ML-based AQP system. ML-AQP offers a complementary approach to that of sampling-based AQP engines. 
whose salient feature is that it develops ML models over (small numbers of) previously executed queries and their answers, instead of developing models or samples over massive base tables. The models are extremely compact and simple, avoiding using complex deep learning networks. Despite this, ML-AQP can provide answers efficiently, offering up to $5$ orders of magnitude speedups in large deployments with small relative errors. Specifically, queries and their answers are transformed into a custom vectorized representation. This representation, allows training ML models that learn patterns to predict future query answers. Also, ML-AQP can bound errors of its answers by employing prediction intervals constructed using Quantile Regression models. In addition, updates to data and changes in query workloads can be handled effectively.
Moreover, it can support any aggregate queries, including \texttt{MIN},\texttt{MAX} (which most AQP systems struggle to address), along with \texttt{GROUP-BYs}. Experiments substantiate that ML-AQP can ensure low errors while introducing dramatic efficiency gains with small memory/storage footprints, and supporting all aggregate functions. ML-AQP, thus, shows a promising method into how ML models can be incorporated for AQP, avoiding the pitfalls of dealing with massive data sets.

\bibliographystyle{abbrv}
\bibliography{sigproc}
\newpage
\appendix
\section{Appendix}
\subsection{Restricting Dimensionality of the Query Representation}
%blinkdb focused on certain most frequent columns
%spatio-temporal analysis is restricted on two columns
%data canopy refer to sdss 55% of queries target non-disticnt columns
As the number of columns (attributes, dimensions) gets larger, 
our representation will be moving towards a high-dimensional space 
causing problems to our underlying ML models. 
One way to tackle this is to use unsupervised dimensionality 
reduction techniques \cite{roweis2000nonlinear}, which will reduce the dimensionality of 
our given query vectors. 
Another, more straightforward way to tackle this is to restrict the number of columns 
we built our ML models over. As examined queries focus on a subset of the original column set \cite{agarwal2013blinkdbaqp,wasay2017data}, allowing us to use a 
heuristic to build ML models only over the most frequently used columns or the columns of the highest 
importance to the analysts. Especially, 
in the case of spatio-temporal data sets the focus is usually on the spatial and temporal 
dimensions to which an analyst applies filters and then examines descriptive statistics over other columns.

\subsection{Sensitivity Analysis}
\label{sec:synth-exp}
\begin{figure}[!htbp]
\begin{center}
\includegraphics[height=5cm,width=7.5cm, keepaspectratio]{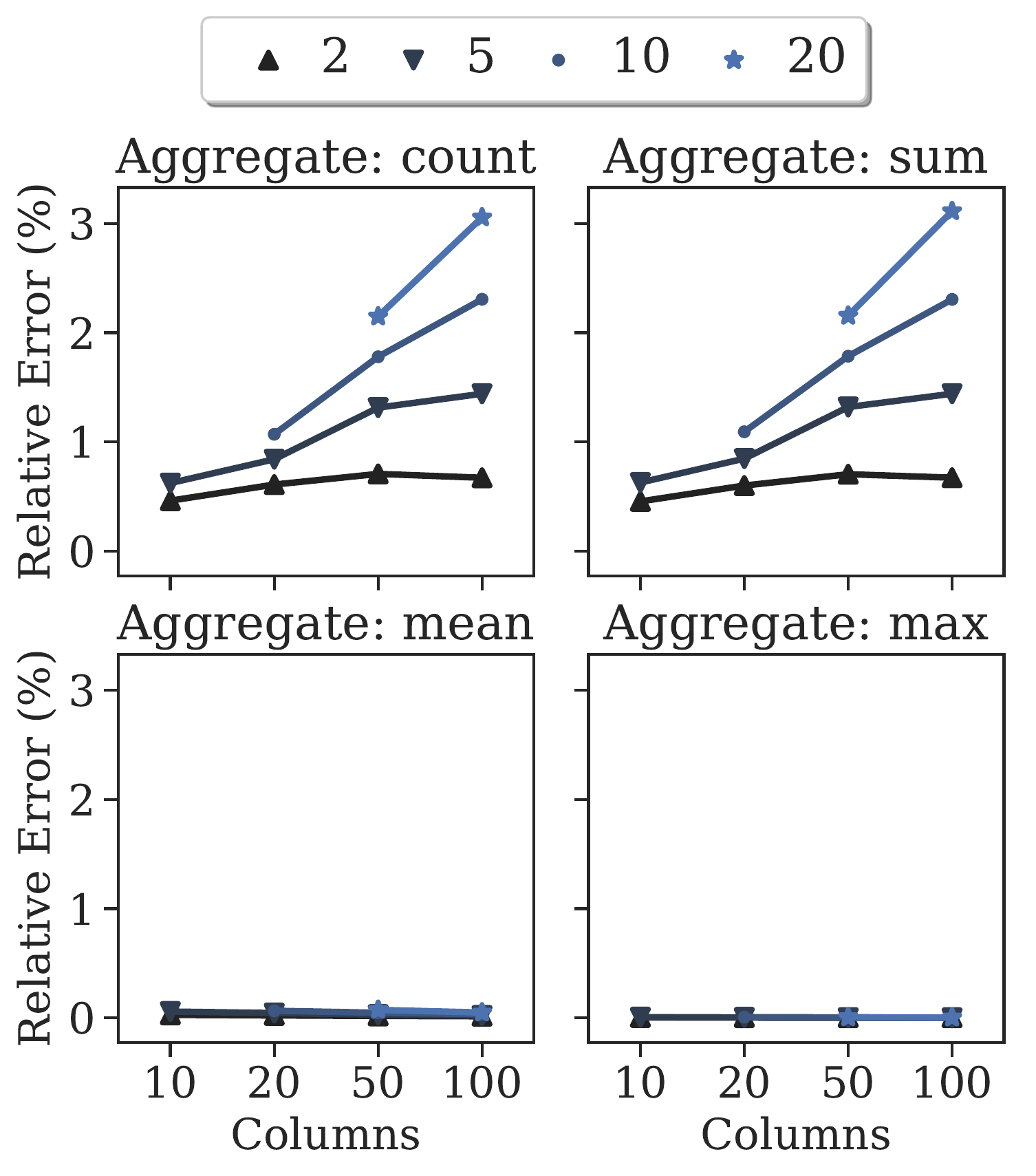}
\caption{Relative Error (y-axis) measured across different aggregates with an increasing number of columns/attributes (x-axis) and a varying number of predicates set randomly.}
\label{fig:rel-error-predicates}
\end{center}
\end{figure}
In this section of our experimental analysis, 
we study a variety of variables contributing to the accuracy of our solution. For these experiments, 
we use the \texttt{synthetic} dataset to control the number of attributes and predicates set. Queries with meta-vectors up to $200$, $\mathbf{m}\in \mathbb{R}^{200}$, are executed over uniform spaces with the number of set predicates up to $50$ \footnote{Note to reviewers: A public Github repository with instructions on how to generate the synthetic datasets as well as automated scripts will be made available. It is merely omitted to adhere to the double-blind constraint.}. 
All predicates are numerical and each query vector is associated with a response. The predicates essentially define range queries over the respective columns. To put this in perspective, real workloads expect a median number of columns selected in a query around $8$ \cite{kandula2016quickraqp} with a more recent estimate reporting that $90\%$ of queries use around $1-6$ \cite{kandula2019experiencesaqp} columns with a maximum reaching $12$. We increase the number of predicates and columns to study the effects on accuracy. We initially train the models on a constant number of queries $10,000$ and vary the number of predicates and columns/attributes. 
In addition, we test different aggregates, \texttt{COUNT}, \texttt{SUM}, \texttt{MEAN}/\texttt{AVG} and \texttt{MAX}, to examine their predictability. In essence, 
ML-AQP is agnostic to what kind of aggregate is being predicted, to ML-AQP an AF applied over an attribute is a response variable to which it tries to identify patterns that can help it minimize the loss $L(y,\hat{y})$.

As can be seen at Figure \ref{fig:rel-error-predicates}, the relative error increases w.r.t. the number of columns and predicates. Although there is not a notable increase ($2\%$), 
we can attribute this to the fact that more queries might be needed to learn a more complex space. 
As the dimensionality of the space (number of columns) increases, the number of predicates increasingly restricts 
the sub-spaces defined by the queries. 
In addition, For \texttt{MEAN} and \texttt{MAX}, 
we do not observe large differences in relative error. 
Closely, examining the workloads we notice that the Coefficient of Variation (CoV), defined by the standard deviation to the mean ration: $\frac{\sigma}{\mu}$ is $0.08$ for the response $y$ of \texttt{MEAN} and $0.01$ for the response \texttt{MAX}. Where the CoV shows the extent of the variability in relation to the mean. A CoV value closer to $1$ indicates high variability. Therefore, ML-AQP might be able to learn their distributions with less queries.  

\begin{figure}[!htbp]
\begin{center}
\includegraphics[height=5cm,width=7.5cm, keepaspectratio]{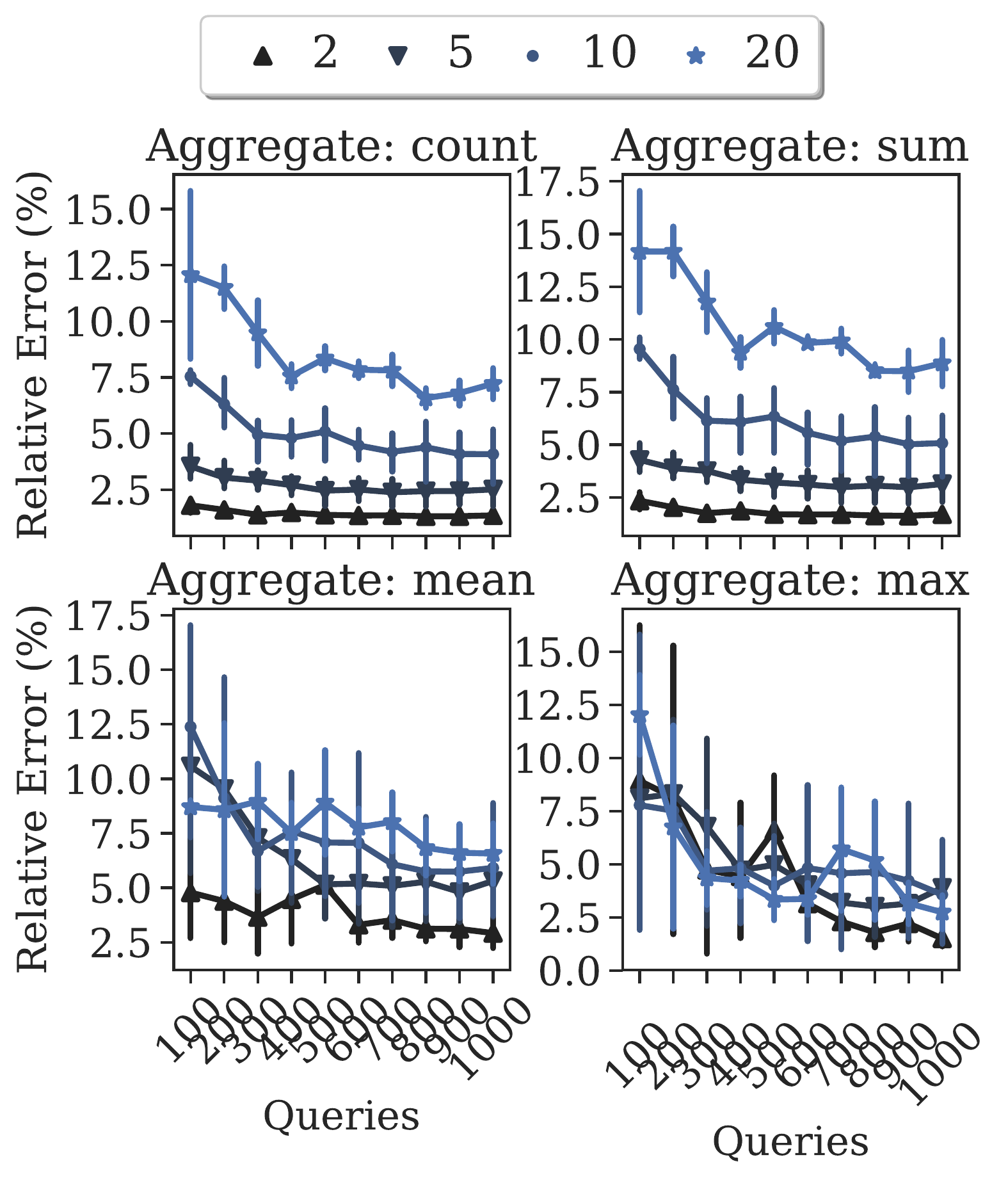}
\caption{Relative Error (y-axis) measured across different aggregates with an increasing number of queries (x-axis) used for training. }
\label{fig:rel-error-queries}
\end{center}
\end{figure}

Figure \ref{fig:rel-error-queries} shows how relative error decreases as the number of queries that a model is trained on increases. For all aggregates, we notice that the effect the number of queries has started to diminish and nothing more can be learned. This provides an approximation to the number of queries needed to model the specific aggregates under this workload. It also denotes that after exceeding a certain number of queries, we should then focus on the complexity of our model to further decrease the relative error. 
In addition, for \texttt{MEAN} and \texttt{MAX}, we see high standard deviation for relative error across settings, especially when the number of queries is small. 
This advises us that as the number of queries is small, 
relative error is largely determined by the complexity of the workload (dimensionality, sparsity). However, as more and more queries are being used to train the models this effect fades away.
\end{document}